\documentclass[manuscript,screen]{acmart}
\usepackage{array}
\AtBeginDocument{%
  }

\copyrightyear{2026}
\acmYear{2026}
\setcopyright{cc}
\setcctype{by}
\acmConference[ASSETS '26]{The 28th International ACM SIGACCESS Conference on Computers and Accessibility}{October 25--28, 2026}{Vila Nova de Gaia, Portugal}
\acmBooktitle{The 28th International ACM SIGACCESS Conference on Computers and Accessibility (ASSETS '26), October 25--28, 2026, Vila Nova de Gaia, Portugal}
\acmDOI{10.1145/3797867.3829032}
\acmISBN{979-8-4007-2521-0/2026/10}

\begin{document}

\title[Are Caption Metrics Broken?]{Are Caption Metrics Broken? Latency, Deaf and Hard of Hearing User Ratings, and Bias across Technologies}


\author{Bernard Thompson}
\email{bernard.thompson@gallaudet.edu}
\orcid{0009-0008-1611-4287}
\affiliation{%
  \institution{Gallaudet University}
  \city{Washington}
  \state{D.C.}
  \country{USA}
}

\author{James Waller}
\email{james.waller@gallaudet.edu}
\orcid{0000-0002-8562-8336}
\affiliation{%
  \institution{Gallaudet University}
  \city{Washington}
  \state{D.C.}
  \country{USA}
}

\author{Luz Fanny Calderon Torres}
\email{luz.fanny.calderon.torres@gallaudet.edu}
\orcid{0009-0009-0416-9560}
\affiliation{%
  \institution{Gallaudet University}
  \city{Washington}
  \state{D.C.}
  \country{USA}
}

\author{Lu Ming}
\email{lu.ming@gallaudet.edu}
\orcid{0009-0008-3771-3684}
\affiliation{%
  \institution{Gallaudet University}
  \city{Washington}
  \state{D.C.}
  \country{USA}
}

\author{Mariana Arroyo Chavez}
\email{marroyoch22@gmail.com}
\orcid{0009-0004-6114-4355}
\affiliation{%
  \institution{Gallaudet University}
  \city{Washington}
  \state{D.C.}
  \country{USA}
}

\author{Dante Conway}
\email{dante.conway@gallaudet.edu}
\orcid{0009-0000-7297-446X}
\affiliation{%
  \institution{Gallaudet University}
  \city{Washington}
  \state{D.C.}
  \country{USA}
}

\author{Raja Kushalnagar}
\email{raja.kushalnagar@gallaudet.edu}
\orcid{0000-0002-0493-413X}
\affiliation{%
  \institution{Gallaudet University}
  \city{Washington}
  \state{D.C.}
  \country{USA}
}

\author{Christian Vogler}
\email{christian.vogler@gallaudet.edu}
\orcid{0000-0003-2590-6880}
\affiliation{%
  \institution{Gallaudet University}
  \city{Washington}
  \state{D.C.}
  \country{USA}
}

\renewcommand{\shortauthors}{Thompson, Waller et al.}

\begin{abstract}
Live captions on TV often contain errors and timing issues, making it hard for deaf and hard-of-hearing (DHH) viewers to follow dialog. It is essential that caption quality metrics reflect the lived DHH TV viewing experience. To this end, we describe a U.S.-based large-scale online survey with 216 validated participants, who provided 302 responses containing a cumulative 4,832 data points. Participants viewed videos drawn from a pool of 70 clips recorded from live TV, and were asked to rate the caption quality and subjective understanding of the content across four conditions: TV captions as originally recorded with up to 7--12 seconds delay, TV captions synchronized with audio, Automatic Speech Recognition (ASR)-generated captions synchronized with audio, and ASR captions with an average two-second delay. All captions were evaluated against the Word Error Rate (WER), Automated Caption Evaluation (ACE2) and Number, Edition and Recognition (NER) metrics. Results show that TV and ASR captions were rated similarly. For TV captions, all three metrics were moderately-to-highly correlated with viewer ratings, but far less so for ASR captions, making them far from technology-neutral. Additionally, caption latencies significantly impact the viewer experience, especially typical 7--12-second TV delays. We discuss the implications for the adoption of caption quality metrics.
\end{abstract}




\maketitle

\section{Introduction}
\label{sec:introduction}

Closed captions, also called subtitles for the deaf and hard of hearing (DHH) in many countries across the world, are a necessary accessibility tool for DHH individuals~\cite{arroyo2024users,butler2019perspectives}, helping them understand spoken content in videos. In the United States, where the study described in this paper took place, the Federal Communications Commission (FCC) is a U.S. government agency that regulates communications, including television, radio, and the internet~\cite{fccabout}. The FCC establishes regulations and guidelines to ensure that captions are accurate, timely, and accessible, emphasizing that clear captions are essential for complying with the law, maintaining viewer satisfaction, and providing equal access to information~\cite{47cfrCaptions}. 

Historically, TV captions have been produced by trained humans, either offline in advance for pre-recorded video, or online in real-time for live TV, such as news, sports, or talkshows. Specifically, under FCC regulations, until the mid-2010s, live captions have been created in one of three ways: (1) human steno captioning as a related discipline of certified court reporting~\cite{Silberstein-Loeb_2009,10.1145/2380116.2380122}, (2) re-speaking by trained voice writers via custom automatic speech recognition (ASR) software~\cite{10.1145/1809777.1809809}, or (3) the electronic newsroom technique, where news anchors read off a pre-written script, which is also fed into the captions~\cite{Silberstein-Loeb_2009}. In the past decade, ASR has increasingly displaced human steno and re-speaking captioning methods~\cite{aliprandi2014automatic}, and been met with a storm of consumer complaints about its quality compared to traditional captioning methods~\cite{forbesbattle}.

At the core of the dispute about ASR captions, and more generally caption quality for live TV is whether they are good enough according to supposedly objective caption quality metrics. TV broadcasters have claimed that ASR captions have an error rate comparable to or better than human captions~\cite{tvnewscheckCaptioningAlready}. One common metric to assess accuracy is the word error rate (WER), which counts every incorrect word, no matter how minor~\cite{nistTools}. Other metrics for captions have been proposed to better reflect the severity of various types of errors, including Automated Caption Evaluation (ACE)~\cite{kafle2017evaluating}, ACE2~\cite{kafle2019predicting}, weighted word error rates (WWER)~\cite{apone2010caption}, and Number, Edition, Recognition (NER)~\cite{romero2015accuracy}; see also Section~\ref{sec:intro-metrics} for further details. 

A key question is whether \emph{any of the metrics in existence} adequately reflect the lived live TV viewing experience of DHH people. Or, in other words, what good is a caption metric if it does not result in better outcomes for the intended target audience? It is also vital to recognize that DHH people are not a monolith. There is substantial evidence that caption needs are highly individual~\cite{lacerda2024royale}; for example depending on how much a person can hear~\cite{armstrong2015diverse}, and that DHH users have widely divergent perspectives on whether any specific captions are good and satisfy their expectations~\cite{wells2022comparing,arroyo2024users,arroyo2024customization}.

While most proposed caption metrics have had some basis in DHH perspectives and industry best practices, there is very little work on validating them through a DHH viewer lens informed by lived experience~\cite{romero2018reception}. ACE~\cite{kafle2017evaluating} and ACE2~\cite{kafle2019predicting} performed a DHH-centric validation, albeit for transcripts in a non-video domain rather than captions. Only one study to date has directly compared metrics against DHH viewer ratings for TV captions~\cite{arroyo2024users} (discussed in Sec.~\ref{sec:related-metrics}), and no comparable validation exists for NER. Since NER has been adopted by government regulators in several countries, including the UK~\cite{ofcom2015measuring} and Canada~\cite{crtc2019}, and has been proposed for adoption in the U.S.~\cite{ner2019fcc}, this validation gap becomes increasingly critical.

The main contribution of this paper is to address the gaps in validating metrics in the U.S. in a DHH-centric way through a large-scale nationwide survey with 216 verified participants via community-based recruiting methods~\cite{uybico2007recruiting}, and 302 responses across 280 distinct broadcast TV video stimuli, resulting in 4,832 video ratings total, in a publicly available dataset (cf. Sec.~\ref{sec:materials}). Beyond the metrics, we also looked at the impact of caption latency, which is very high on U.S. broadcast TV. We additionally examined how the use of ASR may affect metrics and DHH viewer ratings --- a major requirement for regulatory adoption is whether metrics are technology neutral with respect to how the captions were produced. Finally, our survey let participants customize caption appearance freely for font sizes, colors, and positioning, to rule out one-size-fits-all settings as a confounding factor with respect to participant ratings.

\subsection{Research Questions}
\label{sec:rq}
The research questions driving this study were:
\begin{itemize}
\item \textbf{RQ1:} How do caption metrics relate to the DHH viewer experience?
\begin{itemize}
    \item \textbf{RQ1a:} Do results from a large-scale survey replicate findings from prior work?
    \item \textbf{RQ1b:} How does the NER metric relate to the DHH viewer experience?
    \item \textbf{RQ1c:} Do demographic attributes, including age and hearing abilities, affect the DHH viewer experience?
\end{itemize}
\item \textbf{RQ2:} Can metrics be technology-neutral with respect to DHH viewer experience and captioning method?
\item \textbf{RQ3:} How do typical broadcast TV caption latencies affect the DHH viewer experience?
\end{itemize}

NER specifically has a research question of its own under RQ1b, because NER has never explicitly been compared to viewer judgments before. This is in contrast to WER and ACE2, which have been evaluated in past work~\cite{arroyo2024users,wells2022comparing} and just need replication under RQ1a.

\subsection{Caption Quality Metrics in this Study}
\label{sec:intro-metrics}
The metrics assessed in this study were WER, ACE2 and NER. We explain the principles behind each.

\subsubsection{WER}
This metric was developed to assess transcript accuracy, especially for ASR. It weights errors equally, no matter how major or minor, across three categories: (1) deletions, where a word is missing, (2) substitutions, where a word is incorrectly substituted with another, and (3) insertions, where a nonexistent word is added. The WER is the cumulative percentage of all these against the total word count~\cite{nistTools} and ranges from 0 (no errors) to over 100 (if no words are correct). Its key advantage is that it is inexpensive to calculate, because all that is needed is a reference transcript. However, WER does not consider punctuation, nor non-speech information such as background sounds and speaker identification labels.

\subsubsection{ACE2}
This metric was developed to weight transcript errors by their impact, in an ASR context, from a DHH-centric standpoint. The weights are determined by a combination of two factors: (1) how important was the word exhibiting the error; for example missing an article like ``the'' often would be less important than a noun identifying a key concept, and (2) the semantic distance between correct and substituted words. ACE2 employs a neural network-based language model to leverage context for determining word importance~\cite{kafle2019predicting}. Scores typically range between 0 (no errors) and over 1, with no upper bound. The higher the score, the more severe the errors. ACE2 requires aligned transcripts on the sentence level between the reference and hypothesis. At present, this alignment must be performed manually, although once done, the scores can be calculated automatically just like WER. ACE2 does not consider punctuation or non-speech information. 

\subsubsection{NER}
This metric was originally developed for re-speaking live human captions in Europe~\cite{romero2015accuracy}. It classifies errors into two categories: (1) edition, which measures whether paraphrased or otherwise altered captions match the intent of the spoken audio, and (2) recognition, which measures whether information was incorrectly recognized and relayed by the captioner; for example, captioning ``flower'' as ``flour.'' Unlike both WER and ACE2, it considers punctuation and non-speech information, especially speaker labels, as an integral part of the metric. Errors are classified by their severity (no error, minor error, major error) and must be manually assessed by trained evaluators. The NER score is a weighted number between 0 (unusable captions) and 100 (perfect captions), with 98 or higher being classified as good. Like ACE2, NER requires aligned transcripts, and also access to the original video for auditory and visual confirmation. It is by far the most labor-intensive metric to calculate.

\subsection{Positionality Statement}

All but one of the authors identify as DHH. All authors have a background in accessibility adjacent to HCI. All regularly consume captioned videos, although not everyone watches broadcast TV. All but one of the authors are fluent in American Sign Language (ASL). Several of the authors wear hearing aids or cochlear implants and listen to audio while watching video. Several also use spoken English in their everyday communications. All authors use written English, half of them as a second language.

\section{Related Work}

The DHH community comprises individuals with varied communication preferences, language backgrounds, and lived experiences. Existing research suggests considerable variation in captioning preferences, underscoring the need for flexible, inclusive captioning systems~\cite{butler2019perspectives}. Many DHH viewers process captions and visual content at the same time, and this divided attention can increase the impact of presentation rate, caption errors or delays~\cite{lasecki2014,kushalnagar2014collaborative,kushalnagar2014evaluation}. A review of captioning accessibility studies between 2013 and 2023 found that timing and accuracy are the two factors strongly linked to DHH viewer satisfaction~\cite{mcdonnell2024envisioning}.

Generally, related work in the areas of captioning methods, captioning metrics, and captioning appearance shows that many gaps remain. Studies on viewer ratings, as well as head-to-head cross-captioning method comparisons, are especially scarce. This paper contributes toward closing this gap via a large-scale evaluation with a dataset providing full viewer ratings (cf. Sec.~\ref{sec:materials}).


\subsection{Studies on Captioning Methods}
ASR is now used in real-time captioning for live TV broadcasts~\cite{aliprandi2014automatic}. Broadcasters argue that ASR systems produce captions with accuracy that matches or exceeds that of human captions~\cite{perero2018exploring}. However, they face challenges in real-world settings, where factors such as unclear speech, background noise, and varying speaking rates can affect caption accuracy~\cite{perero2018exploring,kuhn2024measuring}. Kuhn et al.~\cite{kuhn2024measuring} examined 11 ASR services and found a difference between performance in controlled tests and accuracy in real-world conditions, especially during live streaming. EvolveCaptions found that adapting ASR in real time for each user can reduce word error rates and improve future captioning performance~\cite{wu2025evolvecaptions}. 

Direct comparisons between different captioning methods are limited. Ruiz-Arroyo et al. compared a respeaking-based system and a fully automatic system on Spanish news broadcasts~\cite{Ruiz-Arroyo2024}, with respeaking producing fewer errors and lower latency. Benchmarking studies have also compared multiple ASR systems with professional human subtitlers using real broadcast content. Lucca and Pierri evaluated four state-of-the-art ASR models on 50 hours of Italian television programming~\cite{lucca2025speech}. They found that although ASR could not fully replace human captioners, it meaningfully improved productivity when used with human review. Related work also investigated how ASR and large language models (LLMs) can be used to improve and double-check steno-generated captions~\cite{wu2024cartgpt,wu2025cartgpt}.

\subsection{Studies on Caption Metrics, Accuracy, and Latency}
\label{sec:related-metrics}
Several variants of the metrics described in Section~\ref{sec:intro-metrics} have been proposed. Weighted Word Error Rate (WWER) accounts for the varying severity of different error types~\cite{apone2010caption}, but was recently shown to relate to the user experience no better than WER~\cite{arroyo2024users}; another extension incorporates punctuation into WER~\cite{datta2020readability}. 

NER~\cite{romero2015accuracy} has received considerable attention in the past decade and has been deployed as a metric in some European countries, as well as Canada. Several evaluation studies of NER on selected U.S. broadcast captions found average scores of over 98 for human-generated captions, which is considered to be good according to NER quality standards~\cite{Romero-Fresco_Fresno_2023,Fresno2024}, while ASR generally scored below that benchmark. However, recent work compared NER scores with Lexi ASR-generated captions and broadcast captions, showing that under the NER metric, ASR caption quality and human-generated caption quality are now very close~\cite{romero2025fit}. Other work investigated the use of fine-tuned LLMs to assess NER automatically on ASR captions, with substantial agreement compared to human evaluators~\cite{romero2024use}.

An important question is whether any of these metrics align with how DHH users experience captions. Arroyo Chavez et al.~\cite{arroyo2024users} addressed this question using a mixed-methods study with 54 DHH participants and 17 hearing participants, comparing live TV captions to 100 percent-accurate reference captions. The study showed that  WER, WWER, ACE, and ACE2 were increasingly strongly correlated with viewer ratings. However, even when captions were highly accurate, they were still seen as a problem, suggesting that accuracy alone does not fully capture the viewing experience. The same team investigated LLM-edited captions, in an attempt to determine whether verbatim captions may be too overwhelming for some users~\cite{arroyo2024customization}, but found that slowing down videos with verbatim captions yielded better benefits. Other work has attempted to model how a human would assess caption quality subjectively~\cite{nam2020modeling}, using machine learning-based classifiers~\cite{nam2023developing}. Another study by Kadoma et al. found that subtitle errors consistently reduced both speaker and content ratings~\cite{10.1145/3772318.3790911}, which highlights the importance of evaluating the viewer experience. 

Caption timing is key for DHH viewers. Szarkowska et al. found that caption speed can have an impact on readability and understanding~\cite{szarkowska2024impact,szarkowska2021effects}. Even small delays between speech and captions can make it harder to read and understand, especially in fast conversations~\cite{burnham1998captions}. Amin et al. found that when captions fall out of sync with audio, viewers struggle to follow the content and match the words when they read with the visuals~\cite{amin2021preferences}. Kuhn et al. found that DHH users preferred verbatim captions with shorter delays compared to edited captions with longer delays, but still appreciated human editing with a small delay~\cite{kuhn2025cart}.

\subsection{Caption Appearance and Personalization}
 Research has explored caption display and personalization in many different settings~\cite{arroyo2024customization,fathallah2024empowering}. Berke et al. showed that DHH participants preferred familiar, closed-caption formatting over alternative appearance styles when using ASR-generated captions in small-group settings~\cite{10.1145/3290607.3312921}, rather than live TV. In contrast, deLacerda-Pataca et al. found that DHH participants preferred customizable styles and better readability and emotional understanding for non-speech information~\cite{lacerda2024royale}. A follow-up study~\cite{de2025cucap} compared how deaf and hard-of-hearing users in North America and South Korea customize such captions, highlighting important cultural differences and the need for flexible design options. 
 
 Feedback from previous studies suggests a strong need for speaker identification features such as color-coded captions or speaker labels~\cite{amin2022preferences}. May et al. found that DHH participants wanted more information about non-speech audio cues than current captions provide~\cite{10.1145/3597638.3608398}. Other recent work investigated haptics to augment non-speech information~\cite{lacerda2025tactile,lacerda2026fuzzy}. Nam et al. found that showing fewer captions during sports broadcasts helped DHH viewers understand the game better without blocking the screen~\cite{10.1145/3806043}, which suggests that genres may affect the viewer experience.

\section{Methods}
\label{sec:methods}

This study was designed to evaluate the quality of live captions in relation to the DHH experience. We employed an online survey that presented captioned video clips across four conditions in a 2x2 factorial within-subjects repeated measures design. One axis of the design was the caption source, which was either the original US broadcast TV captions recorded locally on-site (most of which were likely human-generated, but there is no way to know the captioning method for sure), or ASR-generated captions by a vendor (AppTek) that provides live TV captioning services to U.S. broadcasters. The other axis was latency, which was either zero (i.e., the captions were closely synchronized with the audio), or a form of delay. For TV captions, the delay was the original TV delay, which could go high, typically 7--12 seconds. For the ASR captions, the delay was set to a mean of 2 seconds with a random jitter of up to 50 milliseconds, in an attempt to model typical live ASR latencies that include proper punctuation. The conditions are shown in Table~\ref{tab:conditions}.

\begin{table}[h]
    \centering
    \caption{The factorial design with caption source and caption latency as the axes.}
    \label{tab:conditions}
    \begin{tabular}{|p{0.2\textwidth}|p{0.2\textwidth}|}
    \hline
         \textbf{Source: TV} & \textbf{Source: ASR} \\ \hline
         TV, no delay&  ASR, no delay\\ \hline
         TV, delay as recorded on broadcast (typically 7--12s) &  ASR, mean delay of 2s +/- maximum of 0.05s random jitter\\ \hline
    \end{tabular}
\end{table}

The study design allowed us to compare TV and ASR captions both under ideal latency conditions with no delay, focusing on caption text quality, and under real-world conditions with their respective typical delays. In practice, the choice for captions is often between typical broadcast TV captions with longer and more variable delays, or ASR-generated captions with shorter delays. We restricted the number of conditions to keep participant fatigue in check, choosing these four conditions as the most informative.

The stimuli were a cross-section across U.S. broadcast TV, and totaled 70 distinct video clips, explained further in Section~\ref{sec:materials}. Due to the complexity of this study, we conducted a preliminary focus group to make decisions about the number of stimuli, stimuli duration, default caption settings, caption customization settings, video playback user interface, and even the wording of key measures and accompanying instructions. We refer to this focus group, as applicable, to justify specific aspects of the methods in the following sections. The study received ethics approval, and participants were compensated each time they filled out the survey.

\subsection{Materials}
\label{sec:materials}

We collected video clips recorded from live TV broadcasts by a cable provider in the Washington D.C. area, across 5 genres, as determined by focus group feedback: news, financial, sports, talk shows, and competitions/game shows. These recordings were grabbed by a capture card on a set schedule across multiple channels and resulted in MPEG-2 files with embedded Line-21 captions~\cite{3playmediaWhatClosed}. These captions were extracted and converted to timed text in SRT files. The clips ranged in length from under 30 seconds to 2 minutes, and were divided into buckets of short and long clips, respectively. Long clips were over 60 seconds in duration. Focus group feedback and pilot testing led to the majority of clips being short ones, in order to avoid participant fatigue. On average, each condition featured one long clip and three short ones, hence over the entire survey participants viewed an average of 4 long and 12 short clips. The clips were chosen to be self-contained, such that they could be viewed and understood without having to watch the entire full-length program. While we could not be certain a priori that the duration of the clips would have no statistical effect, the results (cf. Sec.~\ref{sec:results}) do confirm this.

All clips were screened for offensive or controversial content, including hot-button politics and religious topics; such clips were eliminated. Clips where captions featured transmission errors (garbled captions due to broadcast glitches or internet packet loss) were likewise eliminated, with 70 clips remaining. Each clip was prepared under the four conditions described in Table~\ref{tab:conditions}. In total, this resulted in 280 distinct combinations of video clips and timed caption files. All clips and conditions were counterbalanced across participants; see also Section~\ref{sec:instruments}.

For the TV with no delay condition, we used captioning software to adjust the timestamps, ensuring that the captions and audio matched, thereby eliminating any delay. A minimum of two listeners in the research team (typically one DHH and one hearing) independently confirmed the accuracy of the adjusted timestamps. The ASR captions were generated by AppTek, which is a collaborating vendor that serves the U.S. broadcast TV market, with perfectly aligned caption timing. The corresponding delayed ASR captions were created programmatically by varying the timestamps. All video clips had appropriate freeze frames at the end to accommodate the caption delays.

Broadcast TV captions and the ASR captions differed in spelling (such as proper names) and punctuation. An example of both is reflected in this snippet:

\begin{itemize}
    \item \textbf{Broadcast TV}: A FAMILY OF RACCOONS GOT QUITE THE SURPRISE FALLING THROUGH A PORCH CEILING\textbf{.} \textbf{30} MILLION VIEWS ON \textbf{TIKTOK} AND PROBABLY BECAUSE, YOU KNOW, YOU CAN'T JUST WATCH IT ONCE, RIGHT?
    \item \textbf{ASR}: A FAMILY OF RACCOONS GOT QUITE THE SURPRISE\textbf{,} FALLING THROUGH A PORCH CEILING\textbf{,} \textbf{THIRTY} MILLION VIEWS ON \textbf{TICK TOCK,} AND PROBABLY BECAUSE, YOU KNOW, YOU CAN'T JUST WATCH IT ONCE, RIGHT?

\end{itemize}

The dataset with captions and transcripts, the WER, ACE2 and NER evaluations, and the participant-level quality and understanding ratings (keyed by anonymous participant ID, caption source, delay, and video) is available in the supplemental materials accompanying the paper in the ACM DL. Note that participant demographics and open-ended responses are not included because in combination with the ratings they would pose an unacceptable re-identification risk within a relatively small DHH community. They are instead available under managed access via the contact address in the repository. The released materials are sufficient to reproduce all analyses in this paper except those tied to demographics. The most up-to-date version can be found on GitHub\footnote{\url{https://github.com/Gallaudet-University/assets2026-cc}}.

\subsubsection{Survey Instruments}
\label{sec:instruments}
Participants took a series of two Qual\-trics surveys. The first survey collected demographics and also served as a screener to verify eligibility. The second survey, sent via email with links unique to each participant, provided the stimuli. To avoid fatigue, in line with focus group feedback and piloting, each participant was presented with only 16 distinct video clips out of 70, four per condition. The order of conditions and selection of videos was pseudo-randomized based on a cryptographic hash of the participant’s email address, which guaranteed counterbalanced conditions, as well as distinct videos, while maintaining participant privacy. Participants answered questions about quality and understanding after each individual video, as explained in Section~\ref{sec:measures}. They were also given two optional open-ended questions at the end of the survey with written answers: ``Do you have any other comments about the videos you just watched or the captions in them?'' and ``Is there anything else you would like to tell us about live captions on TV? Or anything else that you can think of as important?'' The purpose of these was to give participants an opportunity to feel heard and validated in ways that multiple-choice questions do not support. Participants were allowed to re-take the survey once for additional compensation. Videos were distinct across both iterations of the survey --- no participant ever was presented with the same video twice. The survey typically took 30 minutes.

\subsubsection{Video Playback}
\begin{figure*}[ht]
    \centering
    \includegraphics[width=\textwidth,alt={TV broadcast mentioning the total solar eclipse. Captions are a white font on a black background in all caps. Left: The font is large, and the captions cover most of the bottom of the video. Middle: The font is large, and the captions cover only the left half of the bottom of the video. Right: The font is small , and the captions have many words stretching across the entire width of the video at the bottom.}]{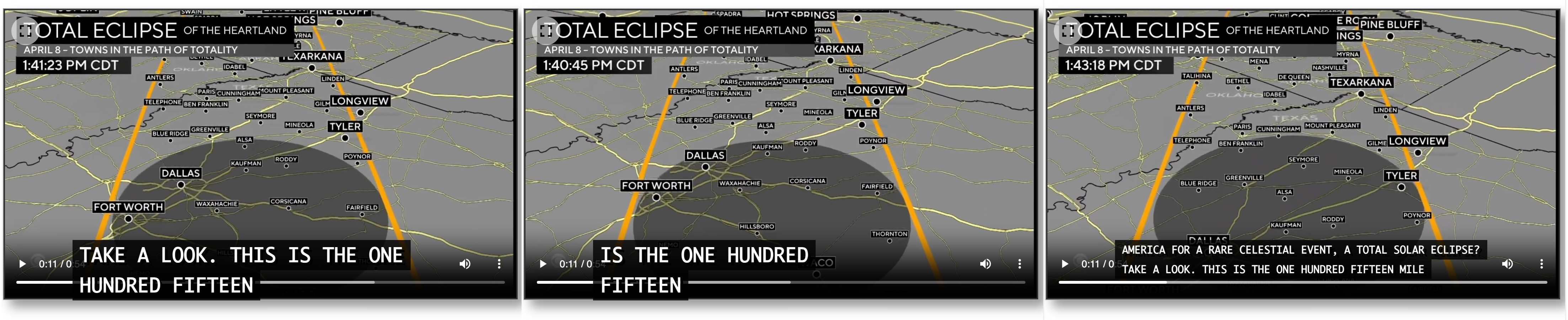}
    \caption{Example of custom video player font size and line settings. Left to right: 150\% caption size with default 32 characters/line setting, 150\% caption size with 20 characters/line, 80\% caption size with 65 characters/line. In fullscreen mode, the size was scaled to maintain the exact proportions to the video.}
    \label{fig:ccsizeline}
    \Description{TV broadcast mentioning the total solar eclipse. Captions are a white font on a black background in all caps. Left: The font is large, and the captions cover most of the bottom of the video. Middle: The font is large, and the captions cover only the left half of the bottom of the video. Right: The font is small , and the captions have many words stretching across the entire width of the video at the bottom.}
\end{figure*}

\begin{figure*}[ht]
    \centering
    \includegraphics[width=\textwidth,alt={TV broadcast mentioning the total solar eclipse. Left (default): Captions are a white font on a black background in all caps, at the bottom of the video and extending partially below. Middle: Captions are at the top of the video. Bottom: Captions are entirely below the video.}]{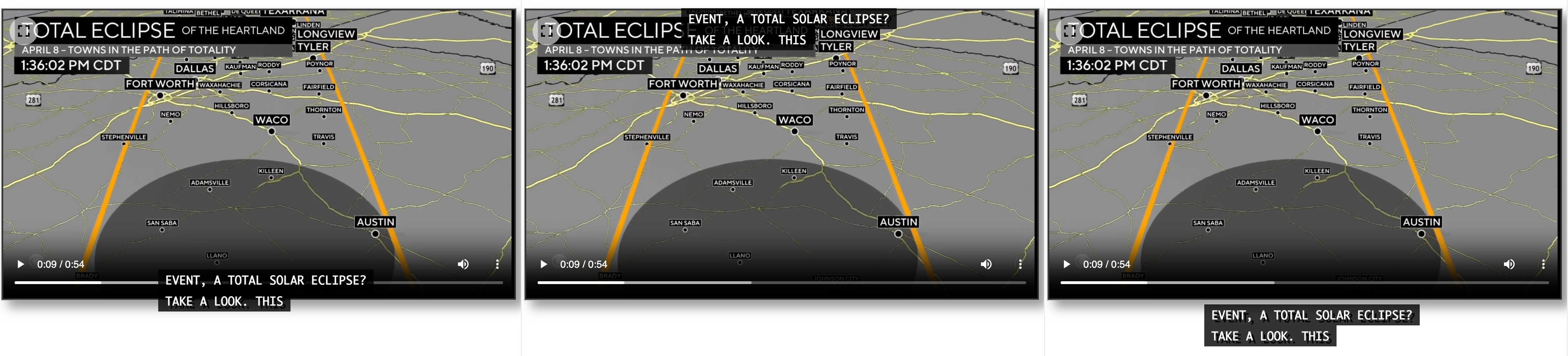}
    \caption{Example of custom video player caption placement settings. Left to right: Default caption placement settings, captions on top, captions below the video. Captions below the video did not block any content. In fullscreen mode, the captions would be moved to the bottom of the full screen such that they were entirely visible.}
    \label{fig:placement}
    \Description{TV broadcast mentioning the total solar eclipse. Left (default): Captions are a white font on a black background in all caps, at the bottom of the video and extending partially below. Middle: Captions are at the top of the video. Bottom: Captions are entirely below the video.}
\end{figure*}

Participants viewed the video clips and accompanying captions with a custom video player embedded into the survey. The player showed captions in the roll-up style characteristic of live TV captions with smooth line transitions in compliance with the FCC rules~\cite{47cfrCaptions}. The user interface was refined with the help of the focus group. It allowed adjustments of font sizes, characters per line (Figure~\ref{fig:ccsizeline}), font colors, background colors, background opacity, as well as captioning positioning anywhere on the video and even below the video (Figure~\ref{fig:placement}); it stored these adjustments, such that settings were persistent on a per-participant basis. It supported windowed and full-screen playback with proper font scaling on major browsers and operating systems, including MacOS and Windows, and iOS and Android on tablets. Participants were given the choice to watch videos with or without audio. They could adjust the audio and caption settings at any time, and replay any video as many times as needed; again, all this was determined by the focus group feedback.

\subsection{Measures}
\label{sec:measures}
After viewing each video, participants answered two questions using a 7-point Likert scale similar to prior related work~\cite{arroyo2024users}: (1) “How would you rate the quality of captions,” and (2) “How much of the content of the videos did you understand?” Two questions were the maximum that the focus group determined to be acceptable for asking after each stimulus without losing the attention of participants. All survey questions were translated into ASL by DHH team members, and participants had the option to view each question individually in ASL or written English.

We calculated metrics for each combination of video clip and TV/ASR captions. A third-party captioning provider generated human-verified offline captions for each clip, which were further verified and corrected by the NER evaluators. We used those as the ground truth for the calculations, described separately for each metric below. The detailed numbers for each video can be found in Appendix~\ref{app:metrics}. In a small number of cases, the SRTs for delayed TV captions differed slightly for those where the captions were synchronized, due to a captioning software editing error, for a total of six affected video stimuli. We have noted these differences, where applicable, in the same appendix. 

\subsubsection{WER}
\label{sec:wer}
We converted the caption files to untimed text transcripts and filtered them to remove punctuation, as well as non-speech information, such as [MUSIC] and speaker name identifications in square brackets, since these parts are beyond the scope of standard WER metrics. The reason is  their inherent variability; for example, [rock music] and [We will rock you by Queen playing] confer similar information, but differ in detail. Resolving these requires human judgment calls, which is counter to the principle of having deterministic evaluation rules. The actual WER scores were calculated via the SCLite tool that is part of the NIST SCTK suite~\cite{nistTools}.  

\subsubsection{NER}
\label{sec:ner}
In line with established practices in other countries, NER was calculated by a team of two certified evaluators, specifically from the Global Alliance of Speech-to-Text Captioning~\cite{speechtotextcaptioningGlobalAlliance}, who were trained for the U.S. captioning market. They cross-checked each others' work and developed consensus on every item. Compared to other countries, the NER scoring was adapted to follow published best practices for broadcast captions in the U.S., as defined by the National Court Reporters Association (NCRA)~\cite{ncra2008}. These go back to a time when steno captioning was considered the gold standard for the U.S. and reflect the expectations of caption viewers from the time before ASR became widespread. Unlike WER and ACE2, the NER evaluation considered punctuation, non-speech information and speaker identification labels.

\subsubsection{ACE2}
\label{sec:ace2}
The scores were calculated based on a language model provided by the original creators of the metric~\cite{kafle2019predicting}. For consistency across NER and ACE2, the alignment of reference and hypothesis units followed the one developed by the NER evaluators. Like for WER, punctuation and non-speech information was stripped prior to calculating the scores. Note that ACE2 scores are calculated on a per-sentence basis. For aggregate ACE2 scores for an entire transcript, we calculated the arithmetic mean across all sentences~\cite{arroyo2024users}. 

\subsection{Participant Recruitment}
Participants were recruited both online and through in-person outreach at events and community gatherings serving DHH individuals. A major challenge with contemporary online surveys is that monetary incentives attract a high number of bad actors, including ineligible individuals and impersonators~\cite{ennis2025exploring}. AI has made this problem even worse by automating survey-taking in ways that are hard to detect after the fact. To avoid compromising the integrity of the survey, a rigorous vetting process was instituted.

Participants who signed up online completed a screening survey that collected demographic information. They were then required to attend a 5-minute video interview using their preferred mode of communication. This interview was used to verify the participants’ demographics, identity (but none of this information was retained), and sign language skills (only if they had indicated in the screening survey that they used ASL), and to confirm eligibility. After verification, the research team provided an explanation of the experiment and emailed a survey link unique to each participant for additional security.

Participants recruited in person at events or conferences, on their own recognizance, could skip the video call verification step and immediately received a unique survey link. They were given the option to complete the survey immediately in the presence of a researcher or on their own time at home.  

In an attempt to generate a demographically balanced sample, we drew on community recruiting approaches~\cite{uybico2007recruiting}. Because our team featured an ethnically diverse membership, many participants were successfully recruited through the networks of the individual researchers. We partnered with DHH community organizations and other community leaders to reach additional people and combined this outreach with travel to events targeting specific demographics, including senior citizens and “grassroots deaf”~\cite{fisher2018deaf}.

\subsection{Procedures}
\label{sec:procedures}

During the video verification call, as well as in-person recruiting, the research team explained what captions are, the role of the FCC, and why captions are important. Participants were also informed that live captions may contain errors and delays. We then explained the survey procedures, sent the survey link by email during the call, and confirmed receipt. Participants then could complete the survey at their convenience. They were also given the opportunity to ask questions.

At the start of the survey, participants were shown a practice video in ASL with captions that allowed participants to adjust the settings in the custom player, as described in Section~\ref{sec:materials}. This practice video contained instructions for adjusting the caption settings and what to expect in the survey. This greatly clarified to participants what they were supposed to do and was based on a focus group suggestion. Participants then were presented with the 16 videos across the 4 conditions, and given the option to answer a few open-ended questions at the end. Participants could pause and resume the survey at any time. To force participants to actually watch the videos, the button to move on to the next question in the survey was hidden until the video playback was complete. 

\subsection{Data Analysis}
\label{sec:analysis}

For analysis of quality ratings, delay, and source, our main modeling approach was a linear mixed model (LMM) with individual responses as the unit of analysis, quality ratings as a numerical outcome and participant and video as crossed random intercepts; source and delay were treated as binary, fixed effects. Delay was simply coded for whether caption delay was present or not --- since delay magnitude would be confounded with source, the interaction term for source and delay was always included. Other additional fixed effects were included in the model where relevant. 

Treating quality and understanding ratings as numerical responses allows for a traditional LMM analysis with interaction contrasts, but assumes equal spacing between levels of the response variable~\cite{liddell2018analyzing}. Assumption checks showed that the distributions of our response variables are broadly appropriate for our LMM analysis~\cite{knief2021violating}, and when we refit the main models as cumulative link mixed models (CLMM)~\cite{burkner2019ordinal} with an ordinal response, our findings were generally unchanged. See Appendix~\ref{app:ordinal} for distribution analysis and ordinal regression results.

Analyses of main effects and their interactions were done with Type III ANOVA using Satterthwaite's degrees of freedom. In some pairwise comparisons, however, it was not feasible to compute exact Satterthwaite \textit{df}, so we used z-tests (essentially assuming infinite \textit{df}). This is an appropriate approximation at this sample size.

We used Holm's method to control for multiple comparisons if several variables were evaluated under a common null hypothesis, including across ANOVA tables and pairwise comparisons. Holm's method is slightly less conservative than Bonferroni: within a family of comparisons, it penalizes the lowest p-values most harshly, with reduced penalty for larger p-values.

Analyses were conducted in R~\cite{r2026language} using the lme4 package~\cite{bates2015package} and ordinal package~\cite{christensen2019ordinal}.

\section{Results}
\label{sec:results}

\subsection{Participant Characteristics}
\label{sec:participants}

\begin{table*}[ht]
\centering
\caption{Participant characteristics and demographic background (n=216).}
\label{tab:participants}
\begin{tabular}{|p{0.34\linewidth}|p{0.6\linewidth}|}
  \hline
  \textbf{Demographic Characteristic} & \textbf{Participant Detail} \\ \hline
  \textbf{Gender} & 65.7\% female, 32.4\% male, 1.9\% other \\ \hline
  \textbf{Self-Identified Hearing Status} & 58.4\% deaf (50.5\% Deaf, 7.9\% deaf), 16.7\% hard of hearing, 6.0\% hearing loss, 3.7\% late-deafened, 13.0\% hearing, rest other \\ \hline
  \textbf{Race/Ethnicity} & 71.8\% White, 15.7\% Hispanic, 10.2\% Asian, 12.5\% Black or African American, 2.8\% American Indian or Alaska Native, 1.4\% Native Hawaiian or Other Pacific Islander. 14.5\% identified as multiracial or multiethnic. \\ \hline
 \textbf{ Education} & 40.3\% master's or higher, 36.6\% bachelor's, 15.7\% some college, 6.0\% high school diploma, <1\% less than high school \\ \hline
  \textbf{Age} & 7.4\% 18--24, 19.9\% 25--34, 18.5\% 35--44, 16.2\% 45--54, 19.0\% 55-64, 19.0\% >65 \\ \hline
  \textbf{Geographic Distribution} & 31 states and U.S. territories \\ \hline
  \textbf{Primary Methods of Communication} & 72.2\% ASL, 14.8\% Signed English (Pidgin or Signed Exact), 48.6\% spoken English, 45.4\% written English (1.4\% exclusively), 5.6\% other languages. Many participants reported more than one language. \\ \hline
  \textbf{Spoken/Signed Languages Irrespective of Written Language Use} & 28.7\% ASL only, 14.4\% ASL and Signed English, 13.9\% spoken English only, 22.7\% ASL and spoken English \\ \hline
  \textbf{Self-Reported English Proficiency} & 87\% high, 10.2\% intermediate \\ \hline
 \textbf{Frequency of Caption Use} & 84.7\% every day, 12.0\% at least twice a week. 92.1\% use live captions \\ \hline
 \textbf{Live Caption Satisfaction} & 51.9\% somewhat or very frustrated, 22.2\% neutral, 22.7\% satisfied \\ \hline
  \textbf{Audio Use with Captions} & 57\% use audio with captions. 27.8\% always have audio on, 20.8\% use audio ``often'' or ``usually.'' \\ \hline
\end{tabular}
\end{table*}

To be eligible, participants had to be 18 years or older, use captions on a regular basis, and have access to an internet-connected computer or tablet. A total of 216 participants completed the 16-item survey, among whom 86 took it twice, with 4,832 rated video clips. The participant demographics and characteristics are summarized in Table~\ref{tab:participants}.

White participants were overrepresented relative to the U.S. population as per the 2020 U.S. census~\cite{censusUSCensus}, while Hispanic participants were slightly underrepresented, and Black participants reflected the census figures. The level of education was higher than the proportions listed in the 2020 census data. The audio use data indicates that while many participants utilized both audio and captions, a significant number relied solely on captions.

In general, background variables related to hearing status, language use, and audio use were correlated. Spearman correlations showed moderate correlations between hearing status and audio use ($\rho$=0.62) and between hearing status and listing a form of English as a primary language ($\rho$=0.47): participants who self-identified as Deaf were less likely to use audio with videos or report English as a primary language or method of communication.

\subsection{Video Clip Caption Metrics}
\begin{figure*}[ht]
    \centering
    \includegraphics[width=\textwidth,alt={TV vs ASR histograms for WER vs ACE2 vs NER. For WER, TV has two modes at 30 and 50. For ACE2, TV has two modes at 0.7 and 0.8. For NER, TV has one mode at 97, with 11 clips above the cutoff of 98. For WER, ASR has one mode at 10. For ACE2, ASR has one mode at 0.45. For NER, ASR has two modes at 93 and 96--97, with 4 clips above the cutoff of 98.}]{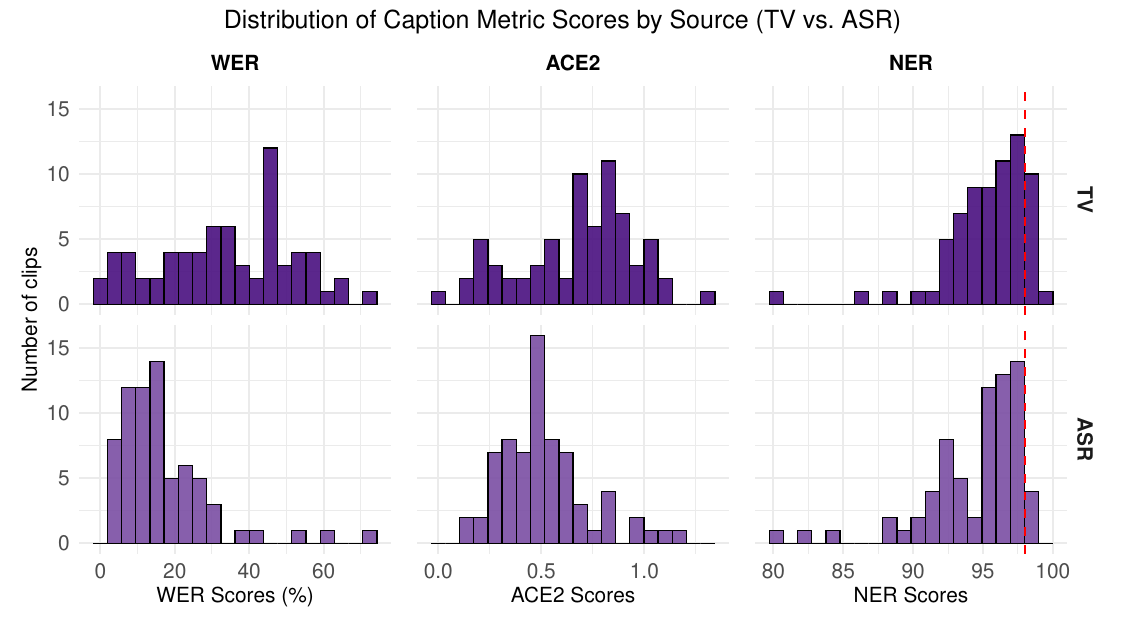}
    \caption{Histogram for the caption quality metrics by transcript (n=70 for each graph). For WER and ACE2, lower is better, while for NER, higher is better. WER and ACE2 are multimodal for TV and more spread out, while they are unimodal for ASR. NER is the opposite with a multimodal distribution for ASR versus a unimodal one for TV. For NER, the cutoff of 98 for acceptable captions is shown, as well.}
    \label{fig:metrics}
    \Description{TV vs ASR histograms for WER vs ACE2 vs NER. For WER, TV has two modes at 30 and 50. For ACE2, TV has two modes at 0.7 and 0.8. For NER, TV has one mode at 97, with 11 clips above the cutoff of 98. For WER, ASR has one mode at 10. For ACE2, ASR has one mode at 0.45. For NER, ASR has two modes at 93 and 96--97, with 4 clips above the cutoff of 98.}
\end{figure*}

 Figure~\ref{fig:metrics} shows the distribution of errors for TV and ASR captions by metric. TV captions exhibited higher error rates and greater variability for WER (mean=33.8, SD=18.0) and ACE2 (mean=0.67, SD=0.28) than ASR captions did for WER (mean=17.2, SD=12.7) and ACE2 (mean=0.52, SD=0.21). For NER, where higher scores are better, the picture was reversed, with TV exhibiting lower error rates and variability (mean=95.28, SD=3.15) than ASR (mean=94.34, SD=3.69). However, only 11 out of 70 TV stimuli met the NER quality threshold of 98, and only 4 out of 70 ASR stimuli did the same.

With respect to ASR, many NER errors centered around the absence of speaker labels, incorrect punctuation, and incorrect non-speech information. ASR captions also sometimes failed to adhere to the NCRA guidelines (cf Sec.~\ref{sec:ner}). Proper names were another problem area, especially with names that are uncommon in the English language. For TV captions, wholesale deletions of content were one of the most common NER errors.

\subsection{Participant Ratings}

We analyzed the total of 4,832 individual ratings for both caption quality and self-reported understanding, spread equally across the four conditions (n=1,208 per condition).

\subsubsection{Caption Quality}
\begin{figure}[ht]
    \centering
    \includegraphics[width=3.2in,alt={Box-and-whisker plots. TV no delay has a median of 5, mean slightly above, box bounds from 4-6. ASR no delay has a median of 5, mean slightly higher than TV, bounds from 4-6. TV with delay has a median of 4, mean slightly below, bounds from 2-5. ASR with delay has a median of 4, mean slightly below, bounds from 4-6. All whiskers span the extremes from 1-7.}]{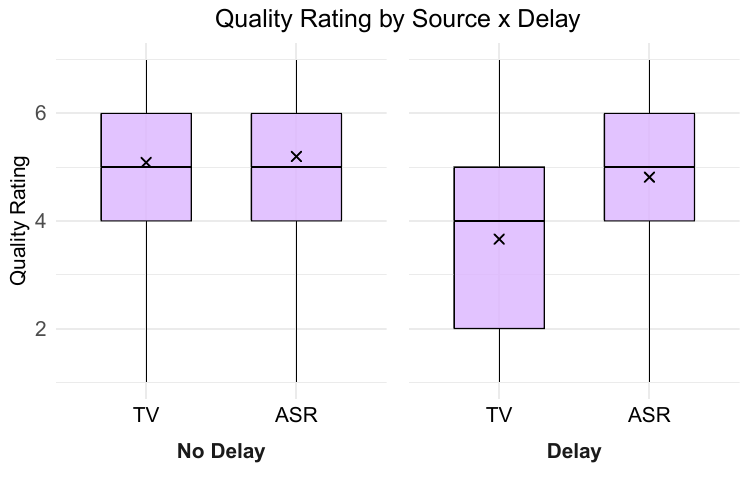}
    \caption{Boxplots of participants' quality ratings across the four conditions. 216 participants (repeated across condition) and 1,208 separate ratings per condition.}
    \label{fig:quality-boxplots}
    \Description{Box-and-whisker plots. TV no delay has a median of 5, mean slightly above, box bounds from 4-6. ASR no delay has a median of 5, mean slightly higher than TV, bounds from 4-6. TV with delay has a median of 4, mean slightly below, bounds from 2-5. ASR with delay has a median of 4, mean slightly below, bounds from 4-6. All whiskers span the extremes from 1-7.}
\end{figure}

On the scale of 1--7, quality ratings for the TV captions presented without delay had a mean of 5.09 (SD=1.50); TV with delay had a mean of 3.66 (SD=1.86); ASR with no delay had a mean of 5.20 (SD=1.41); and ASR with delay had a mean of 4.81 (SD=1.43). Figure~\ref{fig:quality-boxplots} shows the distribution of individual quality ratings across the four conditions.

\begin{figure}[ht]
    \centering
    \includegraphics[width=2.5in,alt={Interaction line plot. In the no delay condition, the TV and ASR mean ratings overlap almost perfectly at a 5.1 mean, with the lines tracing an open jaw toward the delay condition, where the means are 4.7 for ASR and 3.6 for TV. The 95\% confidence intervals are almost identical for all, spanning a range of +/- 0.15.}]{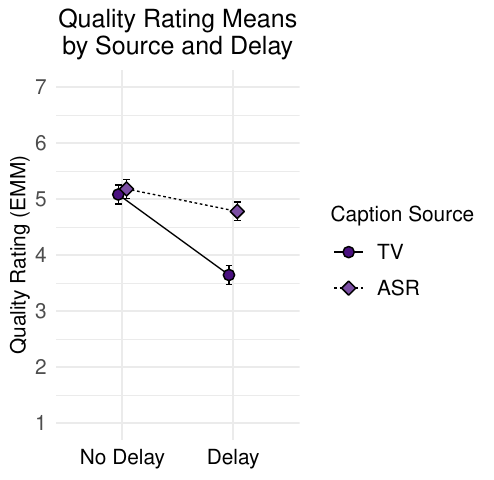}
    \caption{Estimated marginal means for quality ratings across the four conditions. Bars represent 95\% confidence intervals. 216 participants (repeated across condition) and 1,208 separate ratings per condition.}
    \label{fig:quality-emms}
    \Description{Interaction line plot. In the no delay condition, the TV and ASR mean ratings overlap almost perfectly at a 5.1 mean, with the lines tracing an open jaw toward the delay condition, where the means are 4.7 for ASR and 3.6 for TV. The 95\% confidence intervals are almost identical for all, spanning a range of +/- 0.15.}
\end{figure}

To evaluate the effects of delay and caption source on quality ratings, we used an LMM with video and participant random effects (cf Sec.~\ref{sec:analysis}). ANOVA revealed significant main effects of both delay (F(1,4558.9)=623.8, p<0.001) and source (F(1,4557.2)=281.3, p<0.001), and a significant delay x source interaction (F(1,4552.0)=199.5, \linebreak p<0.001): the drop-off in quality ratings is greater for TV captions than for ASR captions when delay is added in. Figure~\ref{fig:quality-emms} shows the marginal means and 95\% confidence intervals. Pairwise tests show that the difference in quality ratings for TV vs. ASR is small (0.10 point difference) and marginally significant (95\% CI $[-0.20,\ 0.003]$, z=-1.90, p=0.057). However, when delay is included the difference between TV vs. ASR ratings is much larger: the average rating for delayed TV captions is over one point lower than delayed ASR captions (-1.14 points, 95\% CI
$[-1.24,\ -1.03]$, z=-21.84, p<0.001). ASR only outperforms TV broadcast captions when comparing delay conditions. This is likely because in the delay conditions the broadcast TV captions have longer delay than the ASR captions (reflecting real-world differences). Under ideal conditions (no latency), TV and ASR captions receive very similar ratings.


\subsubsection{Hearing Status and Quality Perception}

We present an exploratory analysis of how demographic background interacts with the effect of source and delay on quality ratings. We report analyses of only two demographic variables, hearing status and age, although we applied Holm's correction based on all five demographic variables that were explored (hearing status, age, education, audio use, and language).

\begin{figure*}[ht]
    \centering
    \includegraphics[width=5in,alt={Line plot across the deaf, hard of hearing and hearing categories. For TV, the delay penalty has a marked peak at 2.0 for hard of hearing people, with deaf and hearing flanking it at 1.25 each. The confidence intervals are +/- 0.2 for deaf participants, +/- 0.45 for hard of hearing participants, and +/- 0.5 for hearing participants. The ASR delay penalty does not show a peak, instead starting at 0.25 for deaf participants and increasing to 0.5 for both hard of hearing and hearing participants. The ASR confidence intervals are tighter than the ones for TV, at +/- 0.2 deaf, +/- 0.25 hard of hearing, and +/- 0.4 hearing.}]{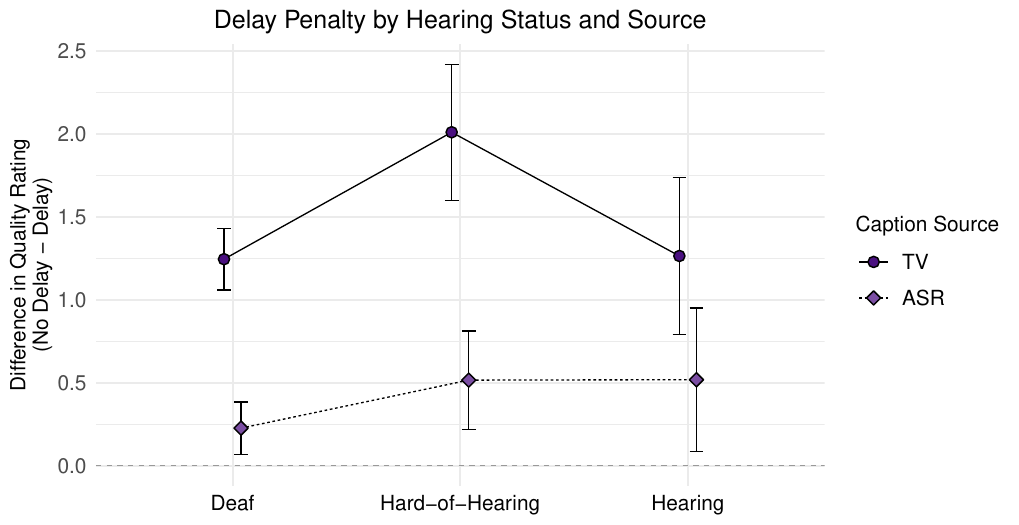}
    \caption{Estimated delay penalties for quality ratings across the four conditions, by hearing status. Delay penalties are the estimated effects of delay (vs. no delay) for a given group x source. Bars represent 95\% confidence intervals.}
    \label{fig:hearing-penalty}
    \Description{Line plot across the deaf, hard of hearing and hearing categories. For TV, the delay penalty has a marked peak at 2.0 for hard of hearing people, with deaf and hearing flanking it at 1.25 each. The confidence intervals are +/- 0.2 for deaf participants, +/- 0.45 for hard of hearing participants, and +/- 0.5 for hearing participants. The ASR delay penalty does not show a peak, instead starting at 0.25 for deaf participants and increasing to 0.5 for both hard of hearing and hearing participants. The ASR confidence intervals are tighter than the ones for TV, at +/- 0.2 deaf, +/- 0.25 hard of hearing, and +/- 0.4 hearing.}
\end{figure*}

We coded hearing status as a factor with three levels: deaf (n=126), hard-of-hearing (n=57; including hearing loss and late-deafened), and hearing (n=28). We excluded data from five participants who did not answer the question or gave a different response. We constructed an LMM with hearing status as a participant-level effect, source and delay as response-level effects, and video and participant as random effects. A 3x2x2 mixed ANOVA showed a significant interaction between hearing status and delay (F(2,4446.1)=22.49, p<0.001) and also hearing status and source (F(2,4448.0)=7.83, \linebreak p=0.004), and a marginally significant three-way interaction between all three (F(2,4442.1)=4.83, p=0.056)\footnote{When re-run as ordinal-response CLMM model, the adjusted p-value was significant at 0.018. See Appendix~\ref{app:ordinal}}. The plot in Figure~\ref{fig:hearing-emms} shows estimated marginal means for all four conditions, for each of the three groups of participants. The source x hearing status interaction appears to be driven by a relatively stronger preference for ASR in hard-of-hearing and hearing participants. Similarly, hearing and hard-of-hearing participants also seem more bothered by delay than deaf participants (delay x hearing status interaction). Figure~\ref{fig:hearing-penalty} shows the `delay effect' for each participant group and caption source; higher delay effects indicate that participants penalized delay more heavily for the group x source combination. Hard-of-hearing participants were especially harsh on delayed TV captions, rating them two points lower than TV captions without delay.

\begin{figure*}[ht]
    \centering
    \includegraphics[width=5in,alt={Series of 3 interaction line plots for deaf, hard of hearing and hearing participants for delay vs no delay. For deaf participants, the lines trace an open jaw from a mean of 5 for both TV and ASR no delay, toward 4.9 ASR and 3.9 TV means with delay. The 95\% confidence intervals are +/- 0.2. For hard of hearing participants, the open jaw is bigger and more steeply sloped, tracing from a mean of 5.2 for TV and ASR no delay, toward 4.75 ASR and 3.1 TV means with delay. The 95\% confidence intervals are bigger, too, at +/- 0.3. For hearing participants, the open jaw is imperfect, and also more steeply sloped but not bigger than the deaf one. It traces a mean of 5.0 for TV/5.2 ASR with no delay toward 4.85 ASR and 3.85 TV means with delay. The 95\% confidence intervals are the largest at +/- 0.4.}]{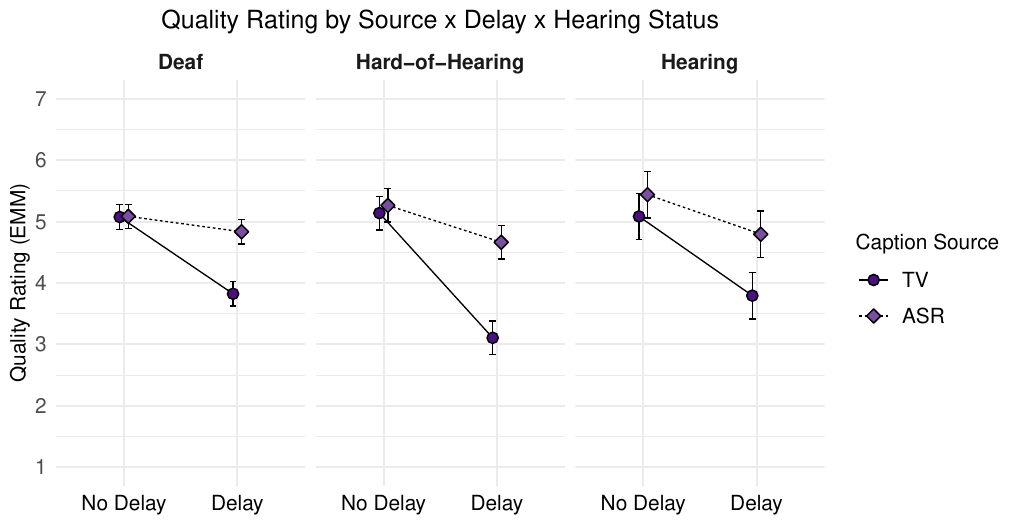}
    \caption{Estimated marginal means for quality ratings across the four conditions, by hearing status. Bars represent 95\% confidence intervals.}
    \label{fig:hearing-emms}
    \Description{Series of 3 interaction line plots for deaf, hard of hearing and hearing participants for delay vs no delay. For deaf participants, the lines trace an open jaw from a mean of 5 for both TV and ASR no delay, toward 4.9 ASR and 3.9 TV means with delay. The 95\% confidence intervals are +/- 0.2. For hard of hearing participants, the open jaw is bigger and more steeply sloped, tracing from a mean of 5.2 for TV and ASR no delay, toward 4.75 ASR and 3.1 TV means with delay. The 95\% confidence intervals are bigger, too, at +/- 0.3. For hearing participants, the open jaw is imperfect, and also more steeply sloped but not bigger than the deaf one. It traces a mean of 5.0 for TV/5.2 ASR with no delay toward 4.85 ASR and 3.85 TV means with delay. The 95\% confidence intervals are the largest at +/- 0.4.}
\end{figure*}

\subsubsection{Age and Quality Perception}
\label{sec:age-quality}
\begin{figure*}[ht]
    \centering
    \includegraphics[width=5in,alt={Line interaction plot for ages under and over 35. Under 35, for no delay, prefers ASR with a margin of 0.3 points over TV. Over 35 has identical ratings for TV and ASR under no delay conditions. Under delay conditions, the difference between TV and ASR is present for both age groups.}]{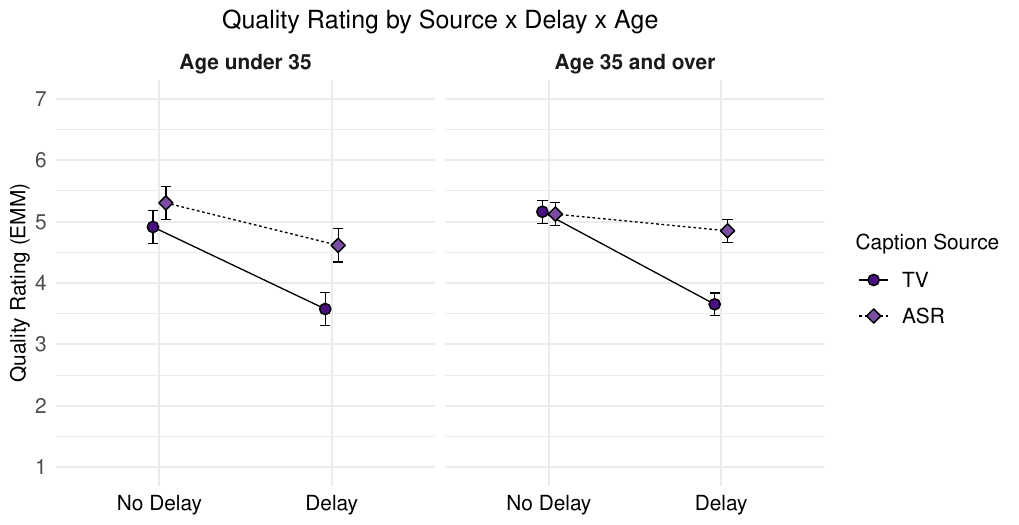}
    \caption{Estimated marginal means for quality ratings across the four conditions, by age (over/under 35 years old). Bars represent 95\% confidence intervals.}
    \label{fig:age-emms}
    \Description{Line interaction plot for ages under and over 35. Under 35, for no delay, prefers ASR with a margin of 0.3 points over TV. Over 35 has identical ratings for TV and ASR under no delay conditions. Under delay conditions, the difference between TV and ASR is present for both age groups.}
\end{figure*}

We hypothesized that older participants are less used to ASR captions and thus rate them lower than younger participants do, who have been exposed to ASR captions since a young age. Specifically, since ASR in the DHH space started in approximately 2010 with the advent of YouTube auto-captions, participants under the age of 35 have experienced the rise of ASR for most of their teen and adult lives, and will have little reference to what broadcast TV captions used to be like in the past. Hence, for this analysis, we treated age as a binary variable, comparing participants under 35 years old (n=59) to participants over 35 years old (n=157). We constructed a linear mixed-effects model with age as a participant-level effect, source and delay as response-level effects, and video and participant as random effects. A 2x2x2 mixed ANOVA showed no significant main effect of age (F(1,207.4)=0.52, p=0.47), nor significant two-way interactions between age and source (F(1,4449.1)=2.48, p=0.46), or age and delay (F(1,4447.8)=2.00, p=0.47). However, there was a significant three-way interaction (F(1,4449.3)=12.16, p=0.004). The plot in Figure~\ref{fig:age-emms} shows that our hypothesis is only partly confirmed. In the no-delay condition, younger participants prefer ASR relative to older participants---but this age difference in ASR preference is less salient in the delay condition.

\subsubsection{Video Effects: Genre and Length}
\begin{figure*}[ht]
    \centering
    \includegraphics[width=5in,alt={Line plots across the 5 genres for TV and ASR, with, and without delay. Without delay, the TV ratings for the financial genre are much lower than for ASR; otherwise the ratings are close to one another. Sports sees a rating dip for both TV and ASR. Under delay, the TV ratings by genre show the same general shape as for no delay, as do the respective ASR ratings. However, they are clearly separated from each other in the delay condition, rather than being close, with TV performing much worse than ASR.}]{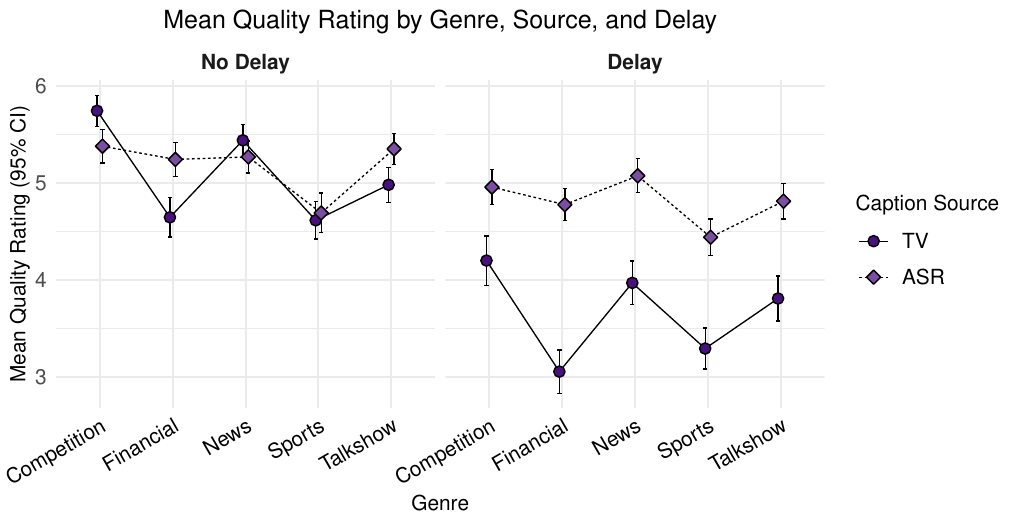}
    \caption{Estimated marginal means for quality ratings across the five genres and four conditions (source x delay). Bars represent 95\% confidence intervals.}
    \label{fig:genre-emms}
    \Description{Line plots across the 5 genres for TV and ASR, with, and without delay. Without delay, the TV ratings for the financial genre are much lower than for ASR; otherwise the ratings are close to one another. Sports sees a rating dip for both TV and ASR. Under delay, the TV ratings by genre show the same general shape as for no delay, as do the respective ASR ratings. However, they are clearly separated from each other in the delay condition, rather than being close, with TV performing much worse than ASR.}
\end{figure*}

We also investigated if quality ratings differed based on properties of the video, including genre and video length. We did not find a significant effect of video length, when including it in the model alongside delay and source (F(1,68.5)=0.49, p=0.48). Short clips (under one minute) were rated about 0.09 points lower than long clips (one to two minutes) in our sample (95\% CI $[-0.33,\ 0.16]$). We did find a significant effect of video genre (F(4,63.7)=9.72, p<0.001). Figure~\ref{fig:genre-emms} shows the estimated marginal means for quality ratings by genre, source, and delay. Visually, Financial and Sports genres have lower ratings than Competition, News, and Talk show. Financial TV ratings are much lower than ASR captions across delay conditions.

Some of these patterns are consistent with gaps seen in caption metrics. For example, the WER for ASR captions of financial videos is only 9.8\% versus 44.6\% for TV captions, and ACE and NER also see large gaps between TV and ASR for financial videos. Other patterns are less easily explained: sports videos receive low quality ratings compared to other genres like talk shows and competitions, but no similar patterns for caption metrics (WER, ACE, or NER) across genres.

\subsubsection{Content Understanding}
Participants rated each video for subjective understanding of the content. TV with no delay had a mean of 5.65 (SD=1.43); TV with delay had a mean of 4.74 (SD=1.79); ASR with no delay had a mean of 5.70 (SD=1.40); and ASR with delay had a mean of 5.49 (SD=1.42). In general the understanding ratings were about 0.5 to 1.0 points higher on average than mean quality ratings. There was a strong, positive correlation between understanding and quality ratings, although not perfect ($\rho$=0.722). 

\begin{figure*}[ht]
    \centering
    \includegraphics[width=5in,alt={Line interaction plots for quality vs understanding. Both show an open jaw, with no delay TV and ASR being nearly identical, and TV relatively dropping off for the delay condition. However, the jaw is smaller for understanding, with visually only about half the drop of ratings. Understanding is 0.5 points higher than ratings for no delay for both TV and ASR. For delay, understanding is 0.5 points higher than ratings for ASR, and 1 point higher for TV. The 95\% confidence intervals are tight for all, at +/- 0.15 points.}]{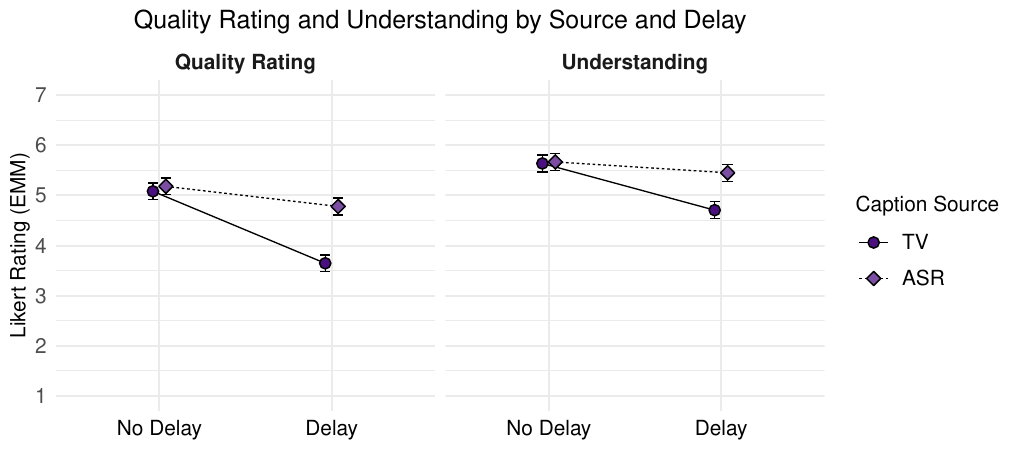}
    \caption{Estimated marginal means for understanding ratings across the four conditions.}
    \label{fig:understanding-emm}
    \Description{Line interaction plots for quality vs understanding. Both show an open jaw, with no delay TV and ASR being nearly identical, and TV relatively dropping off for the delay condition. However, the jaw is smaller for understanding, with visually only about half the drop of ratings. Understanding is 0.5 points higher than ratings for no delay for both TV and ASR. For delay, understanding is 0.5 points higher than ratings for ASR, and 1 point higher for TV. The 95\% confidence intervals are tight for all, at +/- 0.15 points.}
\end{figure*}

We conducted an ANOVA analysis of an LMM with delay and source as fixed effects, but with understanding as the outcome. We found significant main effects of both source (F(1,4553.7)=125.4, p<0.001) and delay (F(1,4555.2)=274.7, p<0.001), with a significant interaction (F(1,4549.1)=107.2, p<0.001). Similar to findings with quality, delay reduced self-reported understanding and the reduction due to delay is larger for TV captions, consistent with TV captions having higher delay overall than ASR captions. Pairwise comparisons showed  no significant differences for TV vs. ASR without delay ($-0.03$ points, 95\% CI $[-0.13,\ 0.07]$, z=-0.62, p=0.54), but with delay included, understanding ratings were significantly lower for TV than ASR by about 0.75 points (95\% CI $[-0.84,\ -0.65]$, z=-15.2, p<0.001). A comparison of marginal means for quality and understanding is shown in Figure~\ref{fig:understanding-emm}. They show that (1) understanding ratings were generally higher than quality ratings and (2) quality ratings were sharply lower for TV delay, but understanding ratings fell less.

\subsection{Participant Ratings and Delay Magnitude}

\begin{figure}[ht]
    \centering
    \includegraphics[width=2.5in,alt={Scatterplot showing ASR and TV delays versus relative rating drops. ASR delays are clustered around 2s, with ratings drops ranging from 1.5 to +0.5. TV delays are mostly clustered between 7s and 12s, with outliers stretching from 0s to over 15s. Rating drops vary widely, but shows a line fit with a downward slope of 1 rating point drop per 7.5s of delay. The confidence interval for the line at the high delay extreme is +/- 0.5 rating drop, and at the low extreme it is 0.35 rating drop.}]{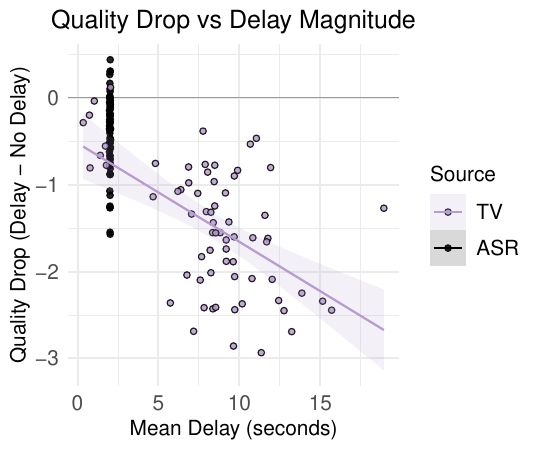}
    \caption{Scatterplot of average delay of a transcript against the quality drop, calculated as the change in mean quality rating from the no-delay condition to the delay condition for the same video x source. Lower (more negative) values indicate larger drops. ASR transcripts (black, n=70) are fixed at average 2 seconds delay; TV transcripts are shown in purple (n=68, two transcripts with negative mean delay excluded). Regression line with 95\% CI is shown for delay-quality relationship for TV.}
    \label{fig:delay-vs-quality}
    \Description{Scatterplot showing ASR and TV delays versus relative rating drops. ASR delays are clustered around 2s, with ratings drops ranging from 1.5 to +0.5. TV delays are mostly clustered between 7s and 12s, with outliers stretching from 0s to over 15s. Rating drops vary widely, but shows a line fit with a downward slope of 1 rating point drop per 7.5s of delay. The confidence interval for the line at the high delay extreme is +/- 0.5 rating drop, and at the low extreme it is 0.35 rating drop.}
\end{figure}

In the delay condition (cf. Table~\ref{tab:conditions}), ASR delay was fixed at 2 seconds while TV delay varied with the original broadcast (M=8.46s, SD=3.62). We investigated how differences in delay magnitude affect quality ratings. For each video we calculated the `quality drop' for both TV and ASR: the change in mean quality rating from the no-delay condition to the delay condition. The scatterplot in Figure~\ref{fig:delay-vs-quality} shows a negative relationship between delay magnitude and change in quality for TV captions: larger delays result in lower, more severe, drops in quality ratings (r=-0.552).

\subsection{Participant Ratings vs Caption Quality Measures}

\begin{figure*}[ht]
    \centering
    \includegraphics[width=5in,alt={Regression scatterplots for all six WER vs ACE2 vs NER against delay vs no delay. Visually, results support line fits for each. The slopes for ASR are similarly flatter for all three metrics, indicating lower correlations; with the direction reversed for NER. The slopes for TV are similarly steeper across the board, indicating higher correlations, again with the direction reversed for NER.}]{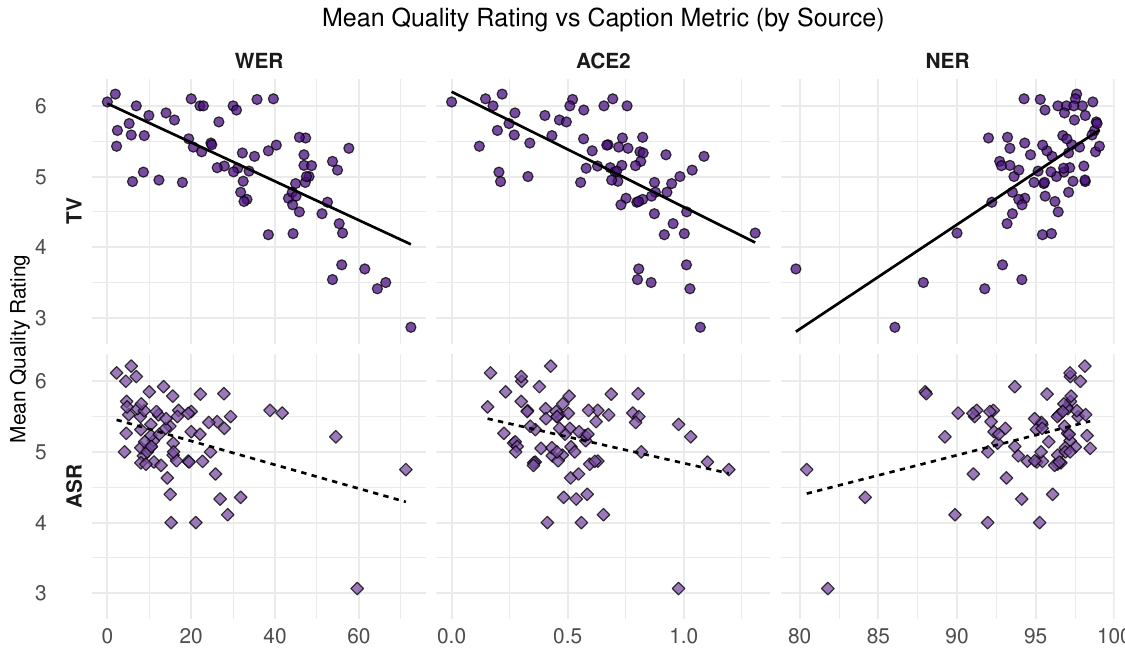}
    \caption{Six scatterplots with regression fits show the relationship between metric and mean quality for each of the three metrics (WER, ACE2, NER) and both TV and ASR (n=70 per condition).}
    \label{fig:metric-scatterplots}
    \Description{Regression scatterplots for all six WER vs ACE2 vs NER against delay vs no delay. Visually, results support line fits for each. The slopes for ASR are similarly flatter for all three metrics, indicating lower correlations; with the direction reversed for NER. The slopes for TV are similarly steeper across the board, indicating higher correlations, again with the direction reversed for NER.}
\end{figure*}

Table~\ref{tab:corr-quality} shows the correlations between caption metrics (WER, ACE2, NER) and mean quality ratings for ASR and TV captions across the 70 videos. Viewer ratings showed strong correlations with all three metrics for TV captions: negative for WER and ACE2, and positive for NER. Correlations for ASR were weaker across the board. We also compared the paired ASR-TV transcripts on different metrics to understand where ASR beat out TV captions, and vice versa. Table~\ref{tab:effect-sizes} shows the paired Cohen's $d$ effect sizes for different ratings and metrics with 95\% CI's. We found low-to-no differences in mean quality or understanding ratings for ASR and TV captions. For WER we found a strong to very strong effect size: TV captions had a much higher WER in our sample videos than ASR captions. We found that ACE2 scores were also higher for TV captions, but only a moderate/strong effect size. NER scores showed a weak to moderate effect in the reverse direction: TV captions had slightly higher NER scores than ASR captions. Figure~\ref{fig:metric-scatterplots} shows the relationship between metric and mean viewer quality ratings for all captions, split by source (TV versus ASR). 

\begin{table}[ht]
\centering
  \caption{Pearson correlations between error metrics (WER, ACE2, NER) and mean quality rating across videos, by caption source. All comparisons for no-delay only. Holm's correction applied to $p$-values within each source; $n = 70$.}
  \label{tab:corr-quality}
  \begin{tabular}{llrr}
    \toprule
    Source & Metric & $r$ & $p_{\text{adj}}$ \\
    \midrule
    TV  & WER  & $-0.687$ & $<.001$ \\
    TV  & ACE2 & $-0.631$ & $<.001$ \\
    TV  & NER  & $\phantom{-}0.647$ & $<.001$ \\
    ASR & WER  & $-0.390$ & $.003$  \\
    ASR & ACE2 & $-0.291$ & $.015$  \\
    ASR & NER  & $\phantom{-}0.383$ & $.003$  \\
    \bottomrule
  \end{tabular}
\end{table}
\begin{table}[ht]
  \centering
  \caption{Paired Cohen's $d$ (TV vs.\ ASR) across outcome measures. Positive values indicate that on average, TV captions scored higher on the given metric than ASR captions.}
  \label{tab:effect-sizes}
  \begin{tabular}{lcc}
    \hline
    Measure & Cohen's $d$ & 95\% CI \\
    \hline
    Quality rating   & $-0.13$ & $[-0.37,\ 0.10]$ \\
    Understanding    & $-0.06$ & $[-0.29,\ 0.18]$ \\
    WER              & $\phantom{-}0.86$ & $[\phantom{-}0.59,\ 1.13]$ \\
    ACE2             & $\phantom{-}0.46$ & $[\phantom{-}0.21,\ 0.70]$ \\
    NER              & $\phantom{-}0.25$ & $[\phantom{-}0.01,\ 0.48]$ \\
    \hline
  \end{tabular}
\end{table}


\subsection{Participant Caption Settings}

\begin{figure}[ht]
    \centering
    \includegraphics[,alt={Chart showing x and y placement of captions superimposed on top of a video. Most are clustered to the left and the bottom of the video, with smaller numbers below the video and at the top of the video. Overall there is significant variability in placement settings.}]{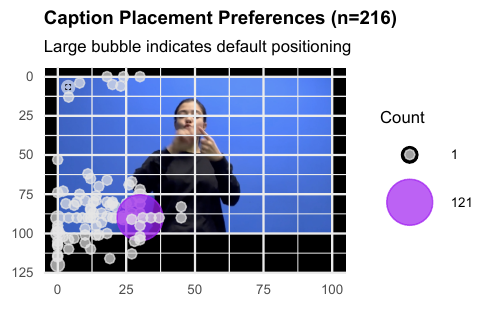}
    \caption{Caption positioning preferences by participants. The majority of placements were at the bottom left to center of the screen, with a sizable number placing them entirely below, and a smaller number placing them at the top.}
    \label{fig:positioning}
    \Description{Chart showing x and y placement of captions superimposed on top of a video. Most are clustered to the left and the bottom of the video, with smaller numbers below the video and at the top of the video. Overall there is significant variability in placement settings.}
\end{figure}

Although caption customization was not a research question, more participants than expected made changes, especially to placement and characters per line. Because of the potential impact on future policymaking, we explicitly report this information here. We found that nearly half of our participants adjusted the caption placement settings. The remainder left them at the default with the captions at the bottom of the video, slightly extending below the video and centered.  The distribution of caption positioning adjustments is shown in Figure~\ref{fig:positioning}.

\begin{figure}[ht]
    \centering
    \includegraphics[,alt={Chart showing how font size and characters per line correlate. Generally, larger fonts have fewer characters per line, and smaller fonts have more. There are long tails both with font size and characters per line. The default is marked at 1.8em font size and 32 characters/line.}]{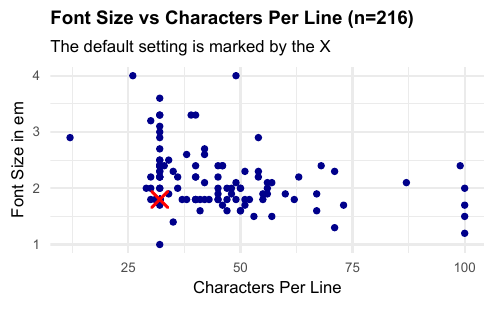}
    \caption{Participants choices of font sizes and characters per line. 1.8 em was the default, as were 32 characters per line. The majority of participants modified at least one of the two, with many preferring more characters per line than the 32 recommended for TV captions.}
    \label{fig:charsperline}
    \Description{Chart showing how font size and characters per line correlate. Generally, larger fonts have fewer characters per line, and smaller fonts have more. There are long tails both with font size and characters per line. The default is marked at 1.8em font size and 32 characters/line.}
\end{figure}

The majority of participants modified the font size and the number of characters per line. Note that participants used computers and tablets to take the survey, and the font sizes were likely dependent on their specific devices and screen sizes. The results are shown in Figure~\ref{fig:charsperline}.

Most participants used the default black background with a white font for the captions, although 10\% used a yellow font on a black background, and 5\% used other color combinations. Likewise, most participants (72\%) used the default opacity of 0.8, with nearly all others using higher values. 

\subsection{Participant Comments}
\label{sec:comments}
In this section we provide some examples from distinct participants to illustrate the participants' open-ended responses and their reactions to the survey stimuli. These are not intended as a full thematic analysis, which is beyond the scope of this study. A related thematic analysis has already been performed in a past mixed-methods captioning study~\cite{arroyo2024users}.

Many participants commented on caption delays, identifying them as the most frustrating aspect of live captioning. \textit{``Delay is most upsetting because you give up trying to read and concentrate on what you hear, which we know is not reliable.'' (P1)}  \textit{``I hate delays --- Within 1 sec is reasonable, but sometimes I see reactions on screen, and I'm not sure why they're reacting or cheering because the captions are behind. I then focus on trying to find out what happened but forget everything else that they talked about.'' (P2)}

Several participants indicated that delay was more disruptive than caption errors.  \textit{``The captions with mistakes were easier to follow than the captions that were delayed, unless a large chunk was missing entirely.'' (P3)} Another participant emphasized \textit{``I would rather have incomplete captions in sync with the spoken words than better captions but lagging behind.'' (P4)}  Other responses further reinforced this sentiment. \textit{``When the captions are not in sync with the speaker's mouth movements, comprehension goes down as it is hard for your brain to put the captions and audio together. [...] I either have to turn the sound off and just read the captions but if the captions are not in sync with the video, there is still a lack of comprehension due to the disconnect between the two.'' (P5)} 

Feedback on the caption adjustment features in the survey was positive. \textit{``The many options offer everyone an opportunity to enjoy watching content with captions in their preferred color, style and also the opportunity to keep adjusting. A very brilliant innovation.'' (P6)} Participants also appreciated being able to move captions. \textit{``I really like the option of adjusting the caption placement to best fit my preferences. For sports events and number-heavy broadcasts such as stock market shows, I can move the captions to the top, and for news and interviews I can move them to the bottom.'' (P7)} However, participants noted that these adjustment features are not commonly found on TV. \textit{``[...] The most important item is the position of the captions because some TV channels have a fixed position that block the graphics.'' (P8)} Speaker identification and non-speech information also were flagged as important: \textit{``Live captions on TV are essential for Deaf and hard-of-hearing viewers. Accuracy, timing, and readability are critical. It's also helpful when captions clearly identify who is speaking and include important sound effects.'' (P9)}

Many participants discussed genre-specific frustrations, especially around sports. \textit{``Out of all of the videos, the sports seemed to be the largest culprits of confusion and bad closed captioning.'' (P10)} News did not fare any better. \textit{``Captioning during live show especially news are notoriously horrendous with a very delayed timing.'' (P11)}

\section{Discussion}
\label{sec:discussion}
Based on our statistical analysis, the results provide strong evidence toward answering the three overarching research questions posed in Section~\ref{sec:rq}. Some of the findings discussed here reinforce results from prior literature, rather than providing new insights. However, they still serve as an important confirmation, due to the much larger sample size and more diverse participant demographics. The much higher statistical power improves both our ability to detect true effects and our confidence that reported effects are real, compared to past work. 

We discuss each of the RQs in turn, and then address the question of whether and how caption metrics should be adopted in the first place. We close out with other notable findings that are not explicitly tied to any of the research questions. 

\subsection{RQ1: Viewer Characteristics and Experiences are Related to Metrics}

With respect to RQ1a, the moderate to strong correlations for TV captions with viewer ratings partially replicate prior work~\cite{arroyo2024users}. Our ACE2 correlation is very close to prior work. In contrast, our WER correlation was as strong as ACE2, but only weak to moderate in prior work. One possible explanation for this difference is that we had a much larger sample size  and that, as a result, the correlations were less sensitive to individual videos. Additionally, as in the same prior work, we found a strong correlation between caption ratings and self-assessed understanding, and one can serve as a proxy for the other. A cautionary note is that self-assessed understanding is different from quiz-based testing of comprehension, but from a DHH-centric HCI standpoint, the viewer's own perceived experience needs to be front and center.

With respect to RQ1b, NER showed as strong a correlation between TV captions and viewer ratings as WER and ACE2 did. To our knowledge, this was the first-of-its-kind study in which NER was assessed against viewer ratings as a proxy for the viewer experience, and it shows that this metric is competitive with the other ones that have been proposed for eventual FCC regulations. However, it does not appear to be superior to WER and ACE, irrespective of caption source. This came as a surprise, given that NER, as evaluated in the U.S., very closely follows established best practices from an age when steno captions reigned supreme in the country. NER penalized missing speaker identification, non-speech information and incorrect punctuation in a way that WER and ACE did not, yet this aspect did not lead to higher correlations. Conversely, NER correlations with ASR captions were only moderate and no higher than ASR with WER, although better than ACE2.

With respect to RQ1c, we saw a clear effect of hearing status on the DHH viewer experience. This showed up in rating differences for delayed captions. Hard of hearing participants penalized delays harshly in their ratings, especially compared to D/deaf participants. This finding mirrors past work that investigated the trade-off between caption accuracy and latency in the UK~\cite{armstrong2015diverse}, where people who listened to audio were much more sensitive to latency. 

Further pursuant to RQ1c, our hypothesis that younger viewers who were exposed to ASR captions for most of their adult lives, would be more tolerant of ASR's limitations (cf. Sec.~\ref{sec:age-quality}) was only partially confirmed: it held for captions synchronized with audio, but not for delayed captions.

In the big picture for RQ1, we see that the DHH viewer experience is reflected in all three metrics that we tested, alongside caption latency. However, this comes with the caveat that TV and ASR captions behaved differently, discussed in the next section.

\subsection{RQ2: Technology-Neutral Metrics Do Not Yet Exist}

An important regulatory principle espoused by DHH consumers is the notion of technology-neutral metrics~\cite{consumerpetition2019}. The expectation is that adherence to caption quality standards should be independent of how the captions were generated. Past caption quality work, especially NER, tacitly made this assumption across steno, re-speaking and ASR captions~\cite{Fresno2024,romero2015accuracy,romero2025fit}. However, the results of our study do not support this principle. 

For a caption metric to be valid for evaluating all caption types, including human-generated captions, ASR-generated captions, and other kinds of captions, it must be (1) unbiased with respect to how the captions were generated, and (2) it must successfully differentiate caption quality for all types of sources as experienced by DHH viewers.  

Point (1) --- unbiased metrics --- failed clearly for WER, and to a lesser extent for ACE2 and even NER. As per Table~\ref{tab:effect-sizes}, WER showed a large gap in scores between the two sources: ASR captions systematically received better WER scores than TV captions. ACE2 scores were also noticeably better for ASR captions. However, \textit{ASR captions were not actually better than TV captions as judged by the viewers}: quality ratings showed negligible differences between TV and ASR captions. This indicates that from a viewer-centric standpoint, WER and ACE2 overestimate the quality of ASR captions relative to broadcast TV ones. NER shows an over-correction in the other direction: it underestimates the quality of ASR captions, penalizing them more harshly than actual human viewers do (although the magnitude of bias is smallest for NER).

The fundamental assumption made under a DHH lens is that viewer experiences, such as ratings, serve as a proxy for caption quality for a grounded DHH-centric caption quality metric. Under this assumption, Point (2) --- differentiating caption quality across both TV and ASR --- also fails. While all three metrics are reasonably successful at distinguishing good TV captions from bad ones with moderate-to-strong correlations with viewer ratings, they are much less successful for distinguishing good ASR captions from bad ones with only weak-to-moderate correlations. If WER, ACE2 and NER were truly technology-neutral with respect to viewer experiences, the correlations for both TV and ASR captions would be expected to be close to one another. 

It should be noted that this finding came as a surprise. WER and ACE2 were specifically developed with ASR in mind, yet ACE2 performed especially poorly at that task under our DHH-centric criteria. Meanwhile, NER did not perform better on the broadcast TV captions than WER and ACE2, despite having been developed with human captioners in mind (which still make up a significant portion of broadcast TV captions). Conversely, it performed comparably poorly on ASR captions.

\subsection{RQ3: Caption Delays Have a Major Impact}

Participants strongly disliked the delays in TV captions as originally broadcast, with a moderate-to-strong correlation between latency and rating drops, mirroring related findings on latency~\cite{szarkowska2021effects}. Beyond the quantitative data, delays also were panned in the participant comments in Section~\ref{sec:comments}. Even the much lower 2s ASR delays resulted in a penalty, especially for hard of hearing and hearing participants who were more likely to listen to audio while watching captions, which tracks with past findings~\cite{burnham1998captions}.

This finding underscores an obvious priority regarding caption quality regulations: addressing delays is a low-hanging fruit, much more so than accuracy metrics. A corollary is that the workflows in caption production and encoding on the TV broadcaster side, all the way to receiving the captions on the viewer side, need to be checked for steps that induce latency. 


\subsection{Metrics, Value Capture and Individual Needs}

Our findings show that all three metrics are reasonably predictive of viewer ratings for broadcast TV captions, but less so for ASR captions. This raises the question: Should metrics be adopted in U.S. regulations, and if so, which ones? There are obvious trade-offs across these three in terms of how expensive they are to calculate, although future advances in AI capabilities have the potential to greatly reduce the cost.

However, there is a much bigger philosophical question lurking in the background: \emph{What is the purpose of adopting caption quality metrics?} The superficial answer is to instill accountability for caption providers and TV broadcasters, especially in light of DHH perceptions that caption quality has declined~\cite{forbesbattle}. Unfortunately, metrics also have the effect of standardizing values~\cite{nguyen2026score}, pp. 197--198: \textit{``[Metrics] reduce diversity and variety. [...] They especially drain the world of diversity of meanings. They reduce our exposure to deeply different modes of valuing, which erodes our reflective control. It's hard to make an informed choice when you don't even know when there was a choice to be made in the first place.''}

There is a clear and present danger that this kind of value capture --- where metrics become an incentive in themselves, rather than DHH-centric needs --- could happen if captioning quality metrics were adopted by regulators. Value capture in turn would disregard the wildly diverse lived experiences and needs of the DHH population. We saw in our survey results that participant demographics have an impact. Beyond that, our caption quality ratings were variable from participant to participant, with many video ratings showing a standard deviation of over 1.5--2 points out of 7, suggesting a highly individualized relationship to caption quality. Other work reinforces this diversity not just for caption quality~\cite{arroyo2024customization,wells2022comparing}, but also for caption appearance on-screen~\cite{may2025choices,lacerda2024royale,de2025cucap,amin2021preferences,amin2022preferences}. We further cannot disregard that captions do not only contribute to access, but that a DHH lens also requires being able to enjoy watching captioned content~\cite{10.1145/3706598.3713238}. Yet, metrics by themselves do not guarantee enjoyment.

A solution to this problem may lie in adopting captioning metrics to track progress toward verbatim captioning and rich non-speech information. The responsibility for customizing caption content and display to individual preferences could then be shifted to the playback devices and software used by DHH viewers. Such an approach looks increasingly feasible as AI, natural language and image processing technologies advance, but would require major new technical standards efforts to expose the necessary information, as noted in a recent study on non-speech information~\cite{may2024rich}.

The above study on exposing rich non-speech information, which then gets customized by the viewer, also suggests a way forward for future captioning metrics. Such metrics could build on the principle of having \emph{as complete a representation as possible} of what is going on in the video. This shifts the focus away from how captioners and viewers interpret video content in an imperfect manner. Although there always will be an element that is subject to interpretation~\cite{zdenek2015reading}, this approach has the potential to greatly reduce the element of subjectivity by the captioner. Such a prospective future metric could assess this information according to correctness and comprehensiveness of the rich representation. The process of translating this representation into viewer-specific captions could consequently be evaluated independently of the rich representation itself. This process could even be benchmarked against more traditional metrics, especially NER, which handles editing and customization better than WER and ACE2. 

\subsection{Caption Quality Trends on Broadcast TV}

While caption quality on broadcast TV in the U.S. was not a research question for this study, we can compare the metrics for our 70 video clips to other work. In that respect, our NER scores were lower than those of studies that specifically attempted to assess caption quality~\cite{Fresno2024,Romero-Fresco_Fresno_2023,romero2025fit}, with fewer of our clips meeting the acceptable NER threshold of 98. This merits asking whether caption quality has declined in the U.S. overall, or whether this is simply an artifact of the broadcasters and TV programs that were available to us where the study was conducted. For instance, VITAC was known as a top-tier human caption provider in a 2020 evaluation~\cite{Romero-Fresco_Fresno_2023} and not available in our video recordings in 2024.

\subsection{Perceptions of ASR vs Reality}

Our findings also raise the question as to whether continued negative perceptions of ASR captions in the DHH communities are still grounded in reality, or whether they are the result of inertia in the face of rapid technological advances. Our study is not the only one to find that ASR captions are increasingly competitive; a recent NER evaluation of ASR captions had a similar finding~\cite{romero2025fit}.

\subsection{Caption Customization and Standards}

The fact that so many participants adjusted the caption settings (Figs.~\ref{fig:positioning} and~\ref{fig:charsperline}) reinforces the importance of caption controls. Of particular note is that half of the participants changed the placement from the defaults. Furthermore, the majority of participants increased the characters per line from the default 32, enshrined in U.S. captioning standards, implying that we need to rethink this number. Although US TVs do not support flexible positioning, and very few streaming providers do, with the notable exception of YouTube, there appears to be a clear need for this, so as to avoid captions blocking important content.

\section{Limitations}

The screening survey and video call requirement created a large amount of friction. While we are confident that we were able to protect data quality, this came at the cost of making participation more difficult and increasing the time and cost per participant. Participants recruited in person at events or conferences were more likely to complete the screening steps compared to those recruited online, but a significant number never followed up with completing the main survey. These results suggest that while in-person recruitment can increase initial engagement, the friction remained a significant barrier to completing the full study. Additionally, in-person recruitment was expensive, making this approach clearly not sustainable. While we were fortunate to have enough funds for carrying out this survey, the challenges for obtaining representative samples of DHH people in the U.S. have become enormous. These challenges are mirrored in cross-disability studies~\cite{goddard2025accessible}, as well.

The participant demographics skewed toward women and self-identified deaf participants, with hard of hearing participants and non-signers being underrepresented; likewise, less-than-college education levels were underrepresented. Thus, our findings may not fully capture the full range of education, literacy, and diversity of the lived DHH experience. All participants were from the U.S., and the results should not be generalized to other countries. 

The study used ASR captions from only one vendor (AppTek), and results may be different for other ASR vendors; note, however, that the point about metrics not being fully technology-neutral still holds even if other vendors were to differ. Additionally, the ASR delay was set to a fixed 2 seconds with jitter, and may not be reflective of the full range of ASR latency in the real world. Participant ratings about quality and understanding may not reflect their enjoyment of using captions, which is an important facet of a DHH lens. We also do not know which of the TV captions in our stimuli were generated by steno, re-speaking and ASR, respectively. This information is generally only known to broadcasters, and not something that we have access to.

Another limitation is that the complexity of the caption workflows in broadcast TV is such that errors can be introduced at various points in the encoder-to-transmission chain after captions have been generated by the vendor. The observed caption quality of our stimuli may be lower than that of some comparable studies for this reason alone. With this said, from a DHH user experience standpoint, it does not matter what the quality of the captions is at production time, but rather what the end user actually sees. 

\section{Future Work}

Future work will have to look closely at how caption latency affects the DHH viewer experience. We found that delays are detrimental, but more information is needed about where the tipping points are for acceptable versus unacceptable delays, both for broadcast TV and ASR captions. In parallel, it is necessary to examine the end-to-end caption workflows among broadcasters to identify and eliminate the major sources of delays.

There is potential for upside in the ACE2 metric through the introduction of better language models to determine word importance and semantic distance, which may result in a metric that is more predictive of user ratings. Another promising avenue of research is to develop AI or machine learning models that attempt to emulate the perceptual DHH viewer experience. While past work has made attempts at this~\cite{nam2020modeling,nam2023developing}, they currently do not generalize sufficiently. With respect to NER, a possible avenue for improvement is to research to what extent the 2008 NCRA guidelines~\cite{ncra2008} underpinning the US NER evaluation model align with contemporary DHH caption viewer experiences. Additionally, NER stands to greatly benefit from research into using AI and LLMs to conduct the evaluation, which would address cost concerns. Similarly, both NER and ACE2 would benefit from AI-assisted mechanisms to align transcripts automatically rather than manually.  

\section{Conclusion}
This work has provided significant evidence toward putting caption quality metrics in relationship with the DHH viewer experience in the U.S., based on a large-scale survey with high statistical power and carefully vetted participants. For TV captions, WER, ACE2 and NER all have similar moderate to high correlations with viewer ratings, but vary in how expensive they are to calculate. 

None of the three are technology-neutral with substantial biases in their behavior with TV versus ASR captions. The difference between mostly human-generated broadcast TV captions and ASR captions appears in the metrics, but not in participant ratings. If a metric is to be adopted in government regulations, care must be taken to avoid value capture and ensure the diverse lived DHH experience does not fall by the wayside. 

TV caption latencies have a major impact on the viewer experience and need to be addressed. Finally, this work provides evidence that customization options for caption placement and characters per line are important for DHH viewers.

\section*{AI Statement}
No AI was used in the writing of this paper. AI was used to double-check the copyediting of the post-review revisions. AI was also used in part for generating the R code for the statistical analysis and fine-tuning the ggplot figure attributes, but vetted by the authors for correctness.


\begin{acks}
The contents of this paper were developed under a grant from the National Institute on Disability, Independent Living, and Rehabilitation Research (NIDILRR grant number 90DPCP0002). NIDILRR is a Center within the Administration for Community Living (ACL), Department of Health and Human Services (HHS). The contents of this paper do not necessarily represent the policy of NIDILRR, ACL, HHS, and you should not assume endorsement by the Federal Government. Norman Williams wrote the custom video player used to present the video and caption stimuli. Jason Durek ported the ACE2 evaluation code to current versions of Tensorflow and Keras. Jennifer Peirson and Christine Ales led the NER evaluation. Joe Merino helped with the graphs for the caption display settings. Pranav Pidathala assisted with caption stimuli review for accuracy and alignment. Emeka Ude assisted in screening participants. AppTek provided the ASR captions based on their technology deployed with broadcasters.
\end{acks}

\bibliographystyle{ACM-Reference-Format}
\bibliography{references}

\appendix
\section{Distribution Checks and CLMM Models for Ordinal Responses}
\label{app:ordinal}

This section examines whether the results from our linear mixed models (LMMs) with ordinal responses are robust. We include assumption checks for our main linear mixed-model analyses and a side-by-side comparison of LMM and CLMM results (cumulative linked mixed models).

\textit{Response distributions.} Quality ratings ranged from 1--7 across every source x delay condition. For each condition, the ceiling and floor contained under 21\% of responses. Skew is generally low-to-moderate (between -0.61 and +0.17 across conditions). Low skew and floor/ceiling effects suggest LMM models yield similar results as ordinal approaches~\cite{knief2021violating, liddell2018analyzing}. Understanding ratings were more compressed at the ceiling (up to 37\% of responses at ceiling), also resulting in more negative skew (between -0.42 and -1.18 across conditions).

\textit{Comparing Linear vs. Ordinal Regression.} We reran three main LMM models as cumulative link mixed models with the same fixed effects: (a) the model for the effect of source and delay on quality ratings, (b) the LMM model for quality ratings with hearing status included (plus source and delay), and (c) the LMM model for the effect of source and delay on \textit{understanding} ratings. All models also have crossed random intercepts for participants and videos. For each model, the table lists the main effects and interaction terms, with the F-statistic and p-value from both the LMM and the CLMM. At significance level $\alpha=.05$ our conclusions for the null hypotheses were unchanged for all terms except the three-way interaction in the models with hearing status, where the adjusted p-value changed from .056 to .018 (values are bolded).

\begin{table}[ht]
  \caption{Quality rating $\sim$ source $\times$ delay. Linear (LMM) versus ordinal (CLMM) regression for main quality rating model. Tests for LMM are re-reported alongside CLMM tests. Both models have the same fixed effects and crossed random intercepts for participants and videos. Both main effects of source and delay, plus their interaction, are significant in both models at $\alpha=.05$.}
  \label{tab:clmm-quality}
  \begin{tabular}{lrrrr}
    \toprule
    Term & LMM $F$ & LMM $p$ & CLMM $F$ & CLMM $p$ \\
    \midrule
    Source & 281.3 & $<.001$ & 230.5 & $<.001$ \\
    Delay & 623.8 & $<.001$ & 547.1 & $<.001$ \\
    Source $\times$ Delay & 199.5 & $<.001$ & 173.7 & $<.001$ \\
    \bottomrule
  \end{tabular}
\end{table}

\begin{table}[ht]
  \caption{Quality rating $\sim$ source $\times$ delay $\times$ hearing status. Linear (LMM) versus ordinal (CLMM) regression for quality rating model with hearing status included. Tests for LMM are re-reported alongside CLMM tests. Both models have the same fixed effects and crossed random intercepts for participants and videos, and p-values for interaction terms use Holm's adjustment as described in the results. The significance of terms are generally identical at $\alpha=.05$ for both LMM and CLMM models, except for the three-way interaction term.}
  \label{tab:clmm-hearing}
  \begin{tabular}{>{\raggedright\arraybackslash}p{2.4cm}rrrr}
    \toprule
    Term & LMM $F$ & LMM $p$ & CLMM $F$ & CLMM $p$ \\
    \midrule
    Source & 214.8 & $<.001$ & 183.4 & $<.001$ \\
    Delay & 479.9 & $<.001$ & 436.2 & $<.001$ \\
    Hearing status & 1.0 & .37 & 0.9 & .40 \\
    Source $\times$ Delay & 121.2 & $<.001$ & 103.3 & $<.001$ \\
    Source $\times$ Hearing status & 7.8 & .004 & 7.0 & .008 \\
    Delay $\times$ Hearing status & 22.5 & $<.001$ & 22.3 & $<.001$ \\
    Source $\times$ Delay $\times$ Hearing status & 4.8 & \textbf{.056} & 6.1 & \textbf{.018} \\
    \bottomrule
  \end{tabular}
\end{table}

\begin{table}[ht]
  \caption{Understanding rating $\sim$ source $\times$ delay. Linear (LMM) versus ordinal (CLMM) regression for understanding rating model. Tests for LMM are re-reported alongside CLMM tests. Both models have the same fixed effects and crossed random intercepts for participants and videos. Both main effects of source and delay, plus their interaction, are significant in both models at $\alpha=.05$.}
  \label{tab:clmm-understanding}
  \begin{tabular}{lrrrr}
    \toprule
    Term & LMM $F$ & LMM $p$ & CLMM $F$ & CLMM $p$ \\
    \midrule
    Source & 125.4 & $<.001$ & 92.0 & $<.001$ \\
    Delay & 274.7 & $<.001$ & 262.0 & $<.001$ \\
    Source $\times$ Delay & 107.2 & $<.001$ & 87.7 & $<.001$ \\
    \bottomrule
  \end{tabular}
\end{table}

\section{Caption Metrics}
\label{app:metrics}

Tables~\ref{tab:metrics1} and~\ref{tab:metrics2} show the caption metrics for all video stimuli.

\begin{table*}[htb]
\caption{Caption metrics for each video. If a TV-related column contains two entries, the metrics were different for the delayed TV captions, as described in Section~\ref{sec:materials}, and are of the format no delay/delay. For WER and ACE2, lower is better; for NER, higher is better. Bolded sets indicate situations where NER scores were over the acceptable threshold of 98.}
\label{tab:metrics1}
\begin{tabular}{|l l|l l l|l l l|}
\hline
VideoID & Genre & TV WER & TV ACE2 & TV NER & ASR WER & ASR ACE2& ASR NER \\ \hline
g01 & Competition & 1.9 & 0.22 & 97.6 & 17 & 0.63& 92.31 \\
g02 & Competition & 20 & 0.15 & 94.26 & 10 & 0.23& 87.98 \\
g03 & Competition & 24.7 & 0.34 & 94.3 & 31.8 & 0.48& 84.16 \\
g04 & Competition & 8.6 & 0.20 & 96.79 & 15.7 & 0.27& 92.58 \\
g05 & Competition & \textbf{0} & \textbf{0.00} & \textbf{98.63} & 10.4 & 0.98 & 95.4 \\
g06 & Competition & 32.2 & 0.83 & 97.09 & 54.5 & 1.03& 89.23 \\
g07 & Competition & 30 & 0.66 & 96.41 & 20 & 0.45& 92.09 \\
g08 & Competition & 30.8 & 0.57 & 95.59 & 26.2 & 0.62& 92.16 \\
g09 & Competition & 39.6 & 0.69 & 97.52 & 41.7 & 0.78& 90.07 \\
g10 & Competition & 35.7 & 0.52 & 95.29 & 29.4 & 0.81& 91.12 \\
g11 & Competition & 47.3/49.5 & 0.74 & 92/92.18 & 27.8 & 0.80& 88.04 \\
g12 & Competition & 38.4 & 0.61 & 95.71 & 38.8 & 0.59& 91.25 \\
g13 & Competition & 47.5 & 0.33 & 96.32 & 71.2 & 1.19& 80.45 \\
g14 & Competition & 22.1/24.7 & 0.51/0.52 & 96.96/96.95 & 24.2 & 0.79& 95.36 \\
f01 & Financial & 48.1 & 0.98 & 93.62 & 4.6 & 0.30& 97.21 \\
f02 & Financial & 44.9 & 0.95 & 95.51 & \textbf{9.8} & \textbf{0.39} & \textbf{98.46} \\
f03 & Financial & 35.1 & 1.09 & 96.04 & 10.4 & 0.51& 97.52 \\
f04 & Financial & 51.2 & 0.87 & 93.52 & 19.3 & 0.63& 94.47 \\
f05 & Financial & 47.2 & 0.87 & 95.56 & 7.9 & 0.45& 97.7 \\
f06 & Financial & \textbf{22.5} & \textbf{0.81} & \textbf{98.8} & \textbf{5.1} & \textbf{0.46}& \textbf{98.17} \\
f07 & Financial & 46.9 & 0.92 & 94.67 & 11.1 & 0.36& 95.07 \\
f08 & Financial & 45.8 & 1.01 & 96.43 & 12.2 & 0.68& 96.47 \\
f09 & Financial & 72.4 & 1.07 & 86.05 & 8.6 & 0.43& 93.89 \\
f10 & Financial & 38.4 & 0.92 & 95.43 & \textbf{5.7} & \textbf{0.43}& \textbf{98.12} \\
f11 & Financial & 20.5 & 0.72 & 97.79 & 15 & 0.58& 96.08 \\
f12 & Financial & 64.4 & 1.03 & 91.76 & 15.6 & 0.51& 97.25 \\
f13 & Financial & 43.2 & 0.75 & 94.31 & 4.5 & 0.23& 97.02 \\
f14 & Financial & 44.1 & 0.93 & 93.52 & 8 & 0.35& 96.6 \\
n01 & News & \textbf{9.9} & \textbf{0.40} & \textbf{98.15} & 9.2 & 0.59& 96.39 \\
n02 & News & 16 & 0.46 & 97.46 & 12.8 & 0.35& 96.2 \\
n03 & News & \textbf{27.9} & \textbf{0.63} & \textbf{98.06} & 4.7 & 0.15& 96.97 \\
n04 & News & 26.2 & 0.68 & 97.16 & 11 & 0.58& 97.18 \\
n05 & News & \textbf{26.6} & \textbf{0.49} & \textbf{98.85} & 27.9 & 0.51& 93.07 \\
n06 & News & 44.3/44.8 & 1.00 & 96 & \textbf{10.4} & \textbf{0.58}& \textbf{98.26} \\
n07 & News & \textbf{2.4} & \textbf{0.20} & \textbf{98.74} & 7.2 & 0.40& 97.03 \\
n08 & News & \textbf{5.7} & \textbf{0.27} & \textbf{98.58} & 8.8 & 0.33& 97.61 \\
n09 & News & 6.9 & 0.18 & 97.97 & 15.7 & 0.48& 96.4 \\
n10 & News & 8.8 & 0.43 & 96.72 & 8 & 0.36& 95.04 \\
n11 & News & \textbf{2.2} & \textbf{0.12} & \textbf{99.05} & 8 & 0.43& 96.88 \\
n12 & News & \textbf{5.2} & \textbf{0.25} & \textbf{98.9} & 8.1 & 0.48& 96.79 \\
\hline
\end{tabular}
\end{table*}

\begin{table*}[htb]
\caption{Caption metrics for each video, continued. See Table~\ref{tab:metrics1} for the detailed legend.}
\label{tab:metrics2}
\begin{tabular}{|l l|l l l|l l l|}
\hline
VideoID & Genre & TV WER & TV ACE2 & TV NER & ASR WER & ASR ACE2& ASR NER \\ \hline
s01 & Sports & 56.1 & 1.31 & 90 & 19.5 & 1.10& 96.5 \\
s02 & Sports & 55.3 & 0.95 & 93.18 & 22.7 & 0.62& 95.43 \\
s03 & Sports & 55.9 & 1.01 & 92.89 & 24.6 & 0.53& 95.34 \\
s04 & Sports & 53.7 & 0.80 & 94.12 & 26.9 & 0.54& 94.11 \\
s05 & Sports & 31.8 & 0.88 & 97.09 & 9.5 & 0.82& 96.89 \\
s06 & Sports & 54.9 & 1.04 & 93.93 & 14.3 & 0.51& 93.14 \\
s07 & Sports & 61.4 & 0.81 & 79.76 & 59.6 & 0.98& 81.79 \\
s08 & Sports & 45 & 0.86 & 95.6 & 28.7 & 0.65& 89.87 \\
s09 & Sports & 17.9 & 0.56 & 94.89 & 15.2 & 0.56& 95.25 \\
s10 & Sports & 14 & 0.27 & 96.83 & 14.9 & 0.34& 96.83 \\
s11 & Sports & 33.3 & 0.82 & 93.92 & 16.7 & 0.45& 96.38 \\
s12 & Sports & 24.9 & 0.67 & 97.29 & 12.4 & 0.49& 97.13 \\
s13 & Sports & 31.1 & 0.58 & 96.77 & 13.9 & 0.47& 95.41 \\
t01 & Talkshow & \textbf{6} & \textbf{0.21} & \textbf{98.15} & 16.5 & 0.46& 94.94 \\
t02 & Talkshow & 52.5 & 0.80/0.82 & 92.21/91.27 & 25.8 & 0.55& 91.04 \\
t03 & Talkshow & 30.1 & 0.70 & 95.97 & 11.7 & 0.33& 96.72 \\
t04 & Talkshow & 66.4 & 0.86 & 87.86 & 8.9 & 0.27& 95.83 \\
t05 & Talkshow & 32.4 & 0.72 & 97.37 & 4.4 & 0.30& 97.84 \\
t06 & Talkshow & \textbf{12.3} & \textbf{0.69} & \textbf{98.13} & 4.1 & 0.45& 97.19 \\
t07 & Talkshow & 40.3/48.8 & 0.67/0.68 & 95.44/95.59 & 14 & 0.41& 94.93 \\
t08 & Talkshow & 53.7/54.1 & 0.81 & 92.71/92.65 & 10.5 & 0.28& 93.67 \\
t09 & Talkshow & 46.9 & 0.73 & 93.36 & 20 & 0.55& 92.35 \\
t10 & Talkshow & 44.2 & 0.73 & 94.13 & 21.1 & 0.41& 91.95 \\
t11 & Talkshow & 32.6 & 0.81 & 96.22 & 6.9 & 0.30& 97.2 \\
t12 & Talkshow & 19.4 & 0.57 & 96.98 & 2.2 & 0.17& 97.18 \\
t13 & Talkshow & 57.6 & 0.76 & 93.36 & 15.8 & 0.28& 91.96 \\
t14 & Talkshow & 48.7 & 0.79 & 92.82 & 19.3 & 0.50& 91.01 \\
t15 & Talkshow & 45.8 & 0.82 & 93.21 & 13.4 & 0.38& 93.67 \\
t16 & Talkshow & 33.8 & 0.71 & 95.15 & 22.1 & 0.59& 92.65 \\
t17 & Talkshow & 22.9 & 0.76 & 97.42 & 22.2 & 0.69& 95.72 \\
\hline
\end{tabular}
\end{table*}

\end{document}